\documentclass[12pt]{article}
\usepackage{amsmath,amsfonts,amssymb,amsthm,cite}
\usepackage{graphicx}
\usepackage{mathptmx}

\title{Density functional theory on phase space}

\author{Philippe Blanchard
\\
Fakult\"at f\"ur Physik, Universit\"at Bielefeld,\\
D-33615 Bielefeld, Germany
\\ \\
Jos\'e M. Gracia-Bond\'ia
\\
Departamento de F\'isica Te\'orica, Universidad de Zaragoza,\\
E--50009 Zaragoza, Spain
\\ \\
Joseph C. V\'arilly
\\
Escuela de Matem\'atica, Universidad de Costa Rica,\\
San Jos\'e 2060, Costa Rica}

\date{Int.\ J. Quant.\ Chem.\ 112, \ 5 February 2012}

\advance\topmargin by -\headheight
\advance\topmargin by -\headsep
\newtheorem{thm}{Theorem}

\newtheorem{cor}[thm]{Corollary}

\DeclareMathOperator{\csch}{csch}     
\DeclareMathOperator{\diag}{diag}     
\DeclareMathOperator{\tr}{tr}         

\newcommand{\vecc}[1]{\mkern-1mu{\vec{\mkern1mu #1}}} 

\newcommand{\A}{\mathcal{A}}        
\renewcommand{\a}{\alpha}           
\renewcommand{\b}{\beta}            
\newcommand{\braket}[2]{\langle#1\mathbin|#2\rangle} 
\newcommand{\D}{\mathcal{D}}        
\newcommand{\del}{\partial}         
\newcommand{\Dl}{\Delta}            
\newcommand{\dl}{\delta}            
\newcommand{\dn}{{\mathord{\downarrow}}} 
\newcommand{\E}{\mathcal{E}}        
\newcommand{\eighth}{\tfrac{1}{8}}  
\newcommand{\Ga}{\Gamma}            
\newcommand{\ga}{\gamma}            
\newcommand{\half}{\tfrac{1}{2}}    
\newcommand{\HH}{\mathbb{H}}        
\newcommand{\hideqed}{\renewcommand{\qed}{}} 
\newcommand{\K}{\mathbb{K}}         
\newcommand{\ket}[1]{|#1\rangle}    
\newcommand{\ketbra}[2]{|#1\rangle\langle#2|} 
\newcommand{\la}{\lambda}           
\newcommand{\om}{\omega}            
\newcommand{\p}{\vecc p}            
\newcommand{\pd}[2]{\frac{\partial#1}{\partial#2}} 
\newcommand{\pp}{\mathfrak{p}}      
\newcommand{\Q}{\mathbb{Q}}         
\newcommand{\q}{\vecc q}            
\newcommand{\quarter}{\tfrac{1}{4}} 
\newcommand{\R}{\mathbb{R}}         
\renewcommand{\r}{\vecc r}          
\newcommand{\rr}{\mathfrak{r}}      
\newcommand{\s}{\vecc s}            
\newcommand{\shalf}{{\scriptstyle\frac{1}{2}}} 
\newcommand{\squarter}{{\scriptstyle\frac{1}{4}}} 
\renewcommand{\th}{\theta}          
\newcommand{\tihalf}{\tfrac{i}{2}}  
\newcommand{\twobytwo}[4]{\begin{pmatrix}#1& #2\\ #3& #4\end{pmatrix}}
\newcommand{\up}{{\mathord{\uparrow}}} 
\newcommand{\vs}{\varsigma}         
\newcommand{\vth}{\vartheta}        
\newcommand{\word}[1]{\quad\mbox{#1}\quad} 
\newcommand{\wt}{\widetilde}        
\newcommand{\x}{\times}             
\newcommand{\z}{\vecc z}            
\newcommand{\7}{\dagger}            
\renewcommand{\.}{\cdot}            

\newcommand{\ee}{\mathrm{ee}}       
\newcommand{\ext}{\mathrm{ext}}     
\newcommand{\gs}{\mathrm{gs}}       
\renewcommand{\H}{\mathrm{H}}       
\newcommand{\h}{\mathrm{h}}         
\newcommand{\HF}{\mathrm{HF}}       
\newcommand{\mini}{\mathrm{min}}    
\newcommand{\TF}{\mathrm{TF}}       

\makeatletter
\def\section{\@startsection{section}{1}{\z@}{-3.5ex plus -1ex minus
 -.2ex}{2.3ex plus .2ex}{\large\bfseries}}
\def\subsection{\@startsection{subsection}{2}{\z@}{-3.25ex plus -1ex
 minus -.2ex}{1.5ex plus .2ex}{\normalsize\bfseries}}
\makeatother

\begin{document}

\maketitle

\begin{abstract}
Forty-five years after the \textit{point de
d\'epart}~\cite{HohenbergK64} of density functional theory, its
applications in chemistry and the study of electronic structures keep
steadily growing. However, the precise form of the energy functional
in terms of the electron density still eludes us -- and possibly will
do so forever~\cite{SuchV09}. In what follows we examine a formulation
in the same spirit with phase space variables. The validity of
Hohenberg--Kohn--Levy-type theorems on phase space is recalled. We
study the representability problem for reduced Wigner functions, and
proceed to analyze properties of the new functional. Along the way,
new results on states in the phase space formalism of quantum
mechanics are established. Natural Wigner orbital theory is developed
in depth, with the final aim of constructing accurate
correlation-exchange functionals on phase space. A new proof of the
overbinding property of the M\"uller functional is given. This exact
theory supplies its home at long last to that illustrious ancestor,
the Thomas--Fermi model.
\end{abstract}


\bigskip

\rightline{\textit{In memory of Jens Peder Dahl, dear friend and
Wigner function's stalwart}\qquad}

\newpage


\section{Introduction}
\label{sec-introit}

\subsection{Conventions and perspective}
\label{ssc:what-it-is}

In this article Hartree atomic units~\cite{Dahl01book} are used. The
operator Hamiltonian~$\HH$ for $N$~fermions involves only one- and
two-body observables. We work under the Born--Oppenheimer regime and
look exclusively at the electronic problem, regarding the potential
due to presence of nuclei as an external one. To fix ideas, consider
the problem of $N$~electrons in an ion of charge~$Z$ in the common
approximation that neglects spin-orbit interaction and weaker
couplings; so that in configuration space
\begin{align*}
\HH &= \sum_{i=1}^{N-1} \sum_{i<j}^N \K(\q_i,\q_j)
:= \sum_{i=1}^{N-1} \sum_{j>i}^N \biggl[
\frac{\h(\q_i) + \h(\q_j)}{N - 1} + \frac{1}{|\q_i - \q_j|} \biggr]
\\
&:= \sum_{i=1}^{N-1} \sum_{j>i}^N \biggl[ \biggl(
- \frac{\Dl_{\q_i}}{2} - \frac{Z}{|\q_i|} - \frac{\Dl_{\q_j}}{2}
- \frac{Z}{|\q_j|} \biggr) \frac{1}{N-1}
+ \frac{1}{|\q_i - \q_j|} \biggr].
\end{align*}
Remember that the set of all $N$-particle density matrices $D_N$
coincides with the set~$\D^N$ of positive hermitian operators of unit
trace on the Hilbert space of antisymmetric $N$-particle functions.
This is a convex set, and its extreme elements are the pure states.
When the system is in the (normalized) pure state~$\ket{\Psi_N}$ the
$N$-particle density matrix is of the form
$D_N = \ketbra{\Psi_N}{\Psi_N}$; then $D_N^2 = D_N$. Given~$D_N$, one
refers to the integral operator with kernel
\begin{equation}
D_n(1,\dots,n;1',\dots,n')
= \int D_N(1,\dots,N; 1',\dots,n',n+1,\dots,N) \,d(n+1)\dots dN
\label{eq:iron-clad} 
\end{equation}
as the reduced $n$-matrix, or simply as the $n$-matrix. We denote by
the same letter a $n$-matrix and its kernel and employ the standard
notation $1 = (\q_1,\vs_1)$, and so on, for the spatial and spin
variables.%
\footnote{In point of rigour, the partial diagonals
in~\eqref{eq:iron-clad} are not generally defined; however, one can
always make sense of the formula by means of the spectral theorem.}
These reduced matrices are still positive operators with trace~1, but
now they form only a proper subset of~$\D^n$. The case $n = 2$ is of
special importance, since the energy of system takes the form
\begin{equation}
E = \binom{N}{2} \tr(\K D_2).
\label{eq:basic-clue} 
\end{equation}
This is an exact linear functional of~$D_2$, and the ground-state
energy would be obtained by minimizing it on the set of 2-matrices.
Therein lies the rub, as the $N$-representability problem for
2-matrices has never been efficiently solved~\cite{ColemanY00}. This
is why the Hohenberg--Kohn--Sham density functional theory (DFT)
program~\cite{HohenbergK64} and its generalizations, initiated by the
proof that the ground state density determines every property of an
electronic system, still enjoy a tremendous success.

\subsection{Purpose and plan of the article}
\label{ssc:why-it-is}

This paper has the aim of sprucing up an untended corner of the garden
of generalized density functional theories. The original DFT program
is beset by our foreordained ignorance of the exact energy functional,
while the reduced density matrix approach is hobbled by the
unsolvability of the $N$-representability problem for two-electron 
densities.%
\footnote{Even before the results in~\cite{SuchV09} the problem has
been regarded as well nigh intractable. Nevertheless, there are the
formal solution by Garrod and Percus \cite{GarrodP64}, still an
appropriate reference for $N$-representability; Ayers' reformulation
of the latter~\cite{Ayers06}; and other recent remarkable progress
both on it and on pending questions of the representability problem
for 1-matrices \cite{ShenviW06,OckoEtAl11,AltunbulakK08,Klyachko09}.
We return to the Ayers method in the next section.}
We push forward here the use of a quasidensity or quasiprobability on
phase space, which is nothing other than the reduced single-body
Wigner function, whose relation to 1-electron space $\rho(\q)$ and
momentum $\pi(\p)$ densities is immediate.

Mathematically, Wigner functions~\cite{Wigner32,Moyal49} are
equivalent to density matrices; and in this sense our approach, dubbed
Wigner density functional theory (WDFT), is equivalent to
\mbox{1-density} matrix functional theory. The beginnings of the
latter go back to~\cite{nonlocalmonger}; a surge of interest ensued,
followed by a period of relative quiescence. More recently, this
approach has seen a vigorous recent development centered around the
notion of natural orbitals. However, WDFT can be worked out
autonomously, calls for a different type of physical intuition, and
readily lends itself to semi-classical treatments. Like one-body
density matrix functional theory, it sits midway between ordinary~DFT
and the Coulson proposal of replacing the wave function by the
2-matrix~\cite{Coulson60}. It stands to contribute to (and to gain
from) both the Kohn and Coulson programs.

Chief among our motivations is the success of the energetic program by
Gill and coworkers -- see for instance~\cite{GillONB03,GillCONB06} --
to attack the problem of correlation energies via the family of Wigner
intracules. Their method is rooted in the fundamental observation by
Rassolov~\cite{Rassolov99} that relative momentum is as important as
relative position in determining interelectronic correlation; and this
naturally calls for a phase space treatment.

\vspace{6pt}

Since we assume from the reader only basic familiarity with Wigner
quasiprobabilities, we do most things from scratch, starting from a
proof of the Rayleigh--Ritz principle in phase space quantum
mechanics. After this, we give the main theorems for energy
functionals based on the single-body Wigner function, in parallel with
standard treatments. There we are brief, since arguments of this type
have become routine.%
\footnote{We consider fermionic systems only in this article. The
parallel formalism for bosons will be considered elsewhere.}

The harmonium ``atom'' is treated exhaustively after that. Phase-space
methods provide a fast lane towards an elegant and complete
description of this system; it also serves as a training ground for
employing the formalism.

True atomic Wigner functions are considered next. Phase-space
eigenfunctions which are not quasidensities are needed for that. We
characterize them for Gaussian basis wave functions; this result seems
to be new. We perform some numerical atomic computations with those
eigenfunctions.

Afterwards we (precisely define and) study \textit{natural Wigner
orbitals}. This is the heart of our subject. They come in handy to
give a simpler proof than that of~\cite{FrankLSS07} of the
``overbinding'' property of the M\"uller 1-density matrix functional
for helium. Again turning to harmonium, we illustrate with them the
workings of exact functionals. The Shull--L\"owdin--Kutzelnigg series
is analytically summed here, for the first time as far as we know.

In Appendix~\ref{app:precursor}, we place the Thomas--Fermi (TF) models
within our framework. The purpose is mainly pedagogical, illustrating
how they fit harmoniously within WDFT. Another appendix elaborates on 
the characterization problem for Wigner functions.

\vspace{6pt}

The fundamental ideas of the present endeavour are found already
in~\cite{ThesisPPm}; they appeared more developed in~\cite{Mimas}.
W{\l}odarz came upon the same idea by the
mid-nineties~\cite{Wlodarz95}. But apparently his paper had an almost
negligible impact. It is perhaps prudent to clarify that our approach
is not quite in the same spirit as the (then recent) work on a phase
space distribution corresponding to a given electronic density~$\rho$,
reported in the texts~\cite{ParrY89,DreizlerG90}. This had been
conjured by Ghosh, Berkowitz and Parr~\cite{GhoshBP84,Berkowitz86} by
means of heuristic ``local thermodynamics'' arguments. The GBP Ansatz
was put to good use~\cite{ParrRG86,GhoshP86,LeeP87} in the calculation
of Compton profiles, exchange energies and corrections to the
Thomas--Fermi--Dirac energy functional. Since the GBP distribution
cannot be a true Wigner quasiprobability, its success should partially
be attributed to the intrinsic strength of the phase space formalism.

There might be an issue of the use of Wigner quasiprobabilities versus
other correspondences between ordinary quantum mechanics and
$c$-numbers on phase space.%
\footnote{We still find most illuminating the treatment of the
relations between phase space representations given by Cahill and
Glauber~\cite{CahillG69} long ago.}
Let it be said that, in the context of our quest, the mathematical
advantages of Wigner functions are overwhelming. It would be difficult
to develop natural orbital theory on phase space without the tracial
property~\cite{Miranda,Apophis} underlying (uniquely) the
Wigner--Moyal correspondence, that allows for an almost verbatim
translation of the spectral resolution for quantum states. Also the
metaplectic invariance of this correspondence apparently is essential
for the reconstruction of the 2-body Wigner function from the 1-body
Wigner function for harmonium in Section~\ref{sec:harm-foretold}. It
lies behind generalized Wigner functions produced by fractional
Fourier transforms~\cite{ChVourdasB99}, as well.

Last, but in no way least: this article seeks to continue and pay
homage to outstanding work of many years by Dahl and Springborg
(see~\cite{DahlS82,SpringborgD87,SpringborgD99} and references given
later) on atomic and molecular Wigner functions, by addressing several
matters of principle.

\section{Classical Hamiltonians and Wigner functions}
\label{sec:Wigner-picture}

The Wigner quasiprobability corresponding to a density matrix $D_N$ is
given by
\begin{align}
& P_N(\q_1,\dots,\q_N;\p_1,\dots,\p_N;
\vs_1,\dots,\vs_N,\vs'_1,\dots,\vs'_N)
\notag \\
&\qquad = \pi^{-3N} \int D_N(\q_1 - \z_1,\vs_1,\dots,\q_N - \z_N,\vs_N;
\q_1 + \z_1,\vs'_1,\dots,\q_N + \z_N,\vs'_N)
\notag \\
&\hspace{4em} \x e^{2i(\p_1\.\z_1 +\cdots+ \p_N\.\z_N)} 
\,d\z_1\dots d\z_N
\notag \\
&\qquad = \pi^{-3N} \int D_N(1',\dots,N';1'',\dots,N'')
e^{i(\p_1\.(\q''_1 - \q'_1) +\cdots+ \p_N\.(\q''_N - \q'_N))}
\notag \\
&\hspace{4em} \x \dl\bigl(\q_1 - \half(\q'_1 + \q''_1)\bigr)
\dots \dl\bigl(\q_N - \half(\q'_N + \q''_N)\bigr)
\,d\q'_1\dots d\q'_N \,d\q''_1\dots d\q''_N.
\label{eq:stoking-trouble} 
\end{align}
This is symmetric under particle exchange. The definition extends to
transitions $\ketbra{\Psi_N}{\Psi'_N}$,
\begin{align}
& P_{\Psi_N\Psi'_N}(\q_1,\dots,\q_N;\p_1,\dots,\p_N;
\vs_1,\dots,\vs_N;\vs'_1,\dots,\vs'_N)
\notag \\
&\qquad = \pi^{-3N} \int
\Psi_N(\q_1 - \z_1,\vs_1,\dots,\q_N - \z_N,\vs_N)
\overline{\Psi'}_N(\q_1 + \z_1,\vs'_1,\dots,\q_N + \z_N,\vs'_N)
\notag \\
&\hspace{4em} \x e^{2i(\p_1\.\z_1 +\cdots+ \p_N\.\z_N)}
\,d\z_1\dots d\z_N.
\label{eq:stoking-more-trouble} 
\end{align}
We can regard Wigner quasiprobabilities as $2N \x 2N$ matrices on spin
space. When there is no risk of confusion the corresponding spinless
quantity, obtained by tracing on the spin variables, is denoted by the
same name. Then hermiticity of~$D_N$ in~\eqref{eq:stoking-trouble}
translates into reality of~$P_N$. The unit trace property translates
into the normalization condition
\begin{equation}
\int P_N(\q_1,\dots,\q_N;\p_1,\dots,\p_N)
\,d\q_1\dots d\q_N \,d\p_1\dots d\p_N = 1.
\label{eq:hidden-menace} 
\end{equation}
Also, Wigner functions are square-summable and bounded continuous. By
interpolation, $P_N \in L^p(\R^{6N})$ for all $2 \leq p \leq \infty$.
Notice that the above integral at any point of phase space is the
expected value of a Grossmann--Royer parity observable
\cite{Grossmann76,Royer77,Dahl82}; those are Stratonovich--Weyl
(de)quantizers on Kirillov coadjoint orbits
\cite{Miranda,Apophis,Oberon}. Since a parity operator cannot have
expectation value greater than~1, we note that
\begin{equation}
|P_N(\q_1,\dots,\q_N;\p_1,\dots,\p_N)| \leq \pi^{-3N},
\label{eq:bounded-joy} 
\end{equation}
with negative values being possible. This bound cannot be reached in
more than one point, and in view of~\eqref{eq:hidden-menace} it is
often remarked that the support of a Wigner function is of bigger
volume than $(h/2)^{3N}$. As a matter of fact, no square-summable
Wigner function can have compact support in all of its variables,
since then both the corresponding wavefunction and its Fourier
transform would have compact support, which is impossible.

The distinctive trait of the phase space formulation is that expected
values are calculated as phase space averages. Hence the classical
observables reenter the quantum-mechanical game. Let $f$ be a real
function of~$\q_1,\dots,\q_N;\p_1,\dots,\p_N$ (a \textit{symbol}) and
let $\Q(f)$ be its quantized operator version (by the Weyl rule). We
work with symbols symmetric in their arguments. Then we get:
\begin{equation*}
\tr\bigl(\Q(f)D_N\bigr) = \int f(\q_1,\dots,\q_N;\p_1,\dots,\p_N)
P_N(\q_1,\dots,\q_N;\p_1,\dots,\p_N)
\,\,d\q_1\dots d\q_N\,d\p_1\dots d\p_N.
\end{equation*}
One must keep in mind that the symbol for the operator $D_N$ is
\textit{not}~$P_N$, but $(2\pi)^{3N} P_N =: W_N$. That is,
$\Q(P_N) = D_N/(2\pi)^{3N}$. A Wigner quasiprobability behaves in
almost every way like a probability density on phase space, except
that in general is not nonnegative everywhere. It is not so easy,
however, to recognize quantum state representatives among all
functions on phase space: see
Appendix~\ref{app:through-a-glass-darkly}.

\subsection{Reduced Wigner functions}
\label{ssc:red-wigner}

Reduced Wigner quasiprobabilities, up to and including the
single-particle quasidensity~$d$ on phase space, are obtained in the
obvious way by integrating successively over groups of 6~variables. In
particular:
\begin{align}
& d_2(\q_1,\q_2;\p_1,\p_2;\vs_1,\vs_2;\vs'_1,\vs'_2) 
\notag \\
&\qquad := \frac{N(N - 1)}{2} \int
P_N(\q_1,\q_2,\q_3\dots,\q_N;\p_1,\p_2,\p_3\dots,\p_N;
\vs_1,\vs_2,\vs_3,\dots,\vs_N;\vs'_1,\vs'_2,\vs_3\dots,\vs_N)
\notag \\
&\hspace{4em}
d\q_3\dots d\q_N\,d\p_3\dots d\p_N\,d\vs_3\dots d\vs_N;
\notag \\
& d(\q;\p;\vs;\vs') \equiv d_1(\q;\p;\vs;\vs') 
:= \frac{2}{N-1} \int d_2(\q,\q_2;\p,\p_2;\vs,\vs_2;\vs',\vs_2)
\,d\q_2\,d\p_2\,d\vs_2.
\label{eq:hace-falta-ser-tonto} 
\end{align}
By partial integration of~$d$ one gets the 1-electron density~$\rho$
and momentum density~$\pi$:
\begin{equation}
\rho(\q) = \int d(\q;\p;\vs;\vs) \,d\p\,d\vs; \qquad
\pi(\p) = \int d(\q;\p;\vs;\vs) \,d\q\,d\vs.
\label{eq:think-positively} 
\end{equation}
Also the form factors are easily expressed in terms of~$d_1$. Since
dequantization and reduction commute, we may also use formulas
analogous to~\eqref{eq:stoking-trouble} for reduced density matrices.
At each order the reduced Wigner distributions contain the same
information as the corresponding reduced density matrices. It is
convenient to have ``inversion formulas'' to recover the latter
matrices from the Wigner functions. It transpires that
\begin{align}
\Ga(1,2;1',2')
&= \int d_2\Bigl( \frac{\q_1 + \q'_1}{2},\frac{\q_2 + \q'_2}{2};
\p_1,\p_2;\vs_1,\vs_2;\vs'_1,\vs'_2 \Bigr)
e^{i(\p_1\.(\q_1 - \q'_1) + \p_2\.(\q_2 - \q'_2))} \,d\p_1\,d\p_2,
\nonumber \\
\ga(1;1') &= \int d\Bigl( \frac{\q + \q'}{2};\p;\vs;\vs' \Bigr)
e^{i\p\.(\q - \q')} \,d\p,
\label{eq:pasodoble} 
\end{align}
where $\Ga = \half N(N-1) D_2$ and $\ga = N D_1$ as usually defined in
quantum chemistry. (Note that $d_2$ corresponds to~$\Ga$ rather than
to~$D_2$ and $d_1$ to~$\ga$ rather than to~$D_1$.)

\vspace{6pt}

One may usefully translate Coleman's theorems~\cite{ColemanY00} on
representable 1-density matrices into properties of~$d$. This was done
by Harriman~\cite{Harriman90}. Let $\D^1_N$ denote the set of
1-matrices representable by~\eqref{eq:iron-clad}; it happens that
$\D^1_N \subsetneq \D^1$. In fact, $d$ is the 1-body Wigner
quasidensity corresponding to an element of~$\D^1_N$ if and only if
\begin{equation}
0 \leq \int W(\q;\p) d(\q;\p) \,d\q \,d\p \leq 2
\label{eq:nesciunt-qui-faciunt} 
\end{equation}
for any $W$ which is the symbol of an arbitrary element of~$\D^1$.
Recall that $d$ may be regarded as a $2 \x 2$ matrix in spin space. We
can write for instance,
$$
d(\q;\p;\vs,\vs') = \begin{pmatrix}
d_{\up\up}(\q;\p) & d_{\up\dn}(\q;\p) \\
d_{\dn\up}(\q;\p) & d_{\dn\dn}(\q;\p) \end{pmatrix};
$$
and $d(\q;\p)$ in~\eqref{eq:nesciunt-qui-faciunt} means the trace of
$d(\q;\p;\vs,\vs')$ -- or its scalar part, in a more cogent
specification of~$d$ according to the behaviour of its components
under rotation. Analogously $d_2$ may be regarded as a $4 \x 4$ matrix
in spin space or as a direct sum of higher tensor representations of
the rotation group. The Ayers trick -- see~\cite{Ayers06} for instance
-- is easily reformulated here: the 1-body Wigner quasidensity belongs
to $\D^1_N$ if and only if
$$
\int h(\q;\p) d(\q;\p) \,d\q \,d\p \geq E_0[h;N]
$$
for every 1-body potential $h(\q;\p)$, where $E_0[h;N]$ denotes the
ground state energy of $N$ fermions under such potential; likewise the
2-body Wigner quasidensity $d_2$ is $N$-representable if and only if
$$
\int h_2(\q_1,\q_2;\p_1,\p_2) d_2(\q_1,\q_2;\p_1,\p_2)
\,d\q_1\,d\p_1 \,d\q_2\,d\p_2 \geq E_0[h_2;N]
$$
for any 2-body Hamiltonian $h_2$. The ``only if'' part is clear, for
otherwise the variational principle would be violated. The ``if'' part
requires the Hahn--Banach theorem.%
\footnote{Thus the axiom of choice.}

The quasidensity corresponding to a Slater determinant (that is, an
extremal element of~$\D^1_N$) is the sum with equal weight 1 of the
contributions of the occupied orbitals~\cite{SpringborgD87}:
$$
d(\q;\p;\vs;\vs') = \sum_{i=1}^N f_i(\q;\p;\vs;\vs'),
$$
with the $f_i$ being (mutually orthogonal) pure-state Wigner
functions. Note, as well, that in this case
\begin{equation}
\int d(\q;\p)^2 \,d\q \,d\p = \frac{2N}{(2\pi)^3}.
\label{eq:sic-transit} 
\end{equation}
In summary, quasidensities look like \textit{mixed-state} Wigner
quasiprobabilities. Typically, the latter are ``less negative'' than
Wigner functions representing pure states.

\subsection{Spectral theorem and variational principle on phase space}
\label{ssc:the-basics}

The spectral theorem for phase space quantum mechanics was given
in~\cite{Triton,Callisto}. Its formulation demands the concept of
Moyal or twisted product~$\x$, indirectly defined by 
$\Q(f \x g) = \Q(f)\Q(g)$. To alleviate notation in what follows we
employ the convention $q = (\q_1,\dots,\q_N)$, and similarly for the
other variables. The twisted product is
$$
f \x g(q;p) = \int f(q';p') g(q'';p'')
\exp[2i(qp' - q'p + q'p'' - q''p' + q''p - qp'')]\,
\frac{dq'\,dp'\,dq''\,dp''}{\pi^{6N}},
$$
with the properties:
\begin{equation}
\int f \x g(q;p) \,dq\,dp = \int f(q;p) g(q;p) \,dq\,dp
= \int g \x f(q;p) \,dq\,dp; \quad
\overline{f \x g} = \overline g \x \overline f.
\label{eq:rollo-habitual} 
\end{equation}

\begin{thm} 
\label{th:matrix-basis}
Assume for simplicity a purely discrete nondegenerate energy spectrum
$E_0 < E_1 <\cdots< E_n <\cdots$. Then:
\begin{itemize}
\item
The solutions $\Ga_{nm}$, for $n,m = 0,1,2,\dots$, of the simultaneous
eigenvalue equations:
\begin{equation}
H \x \Ga_{nm} = E_n\,\Ga_{nm}, \qquad \Ga_{nm} \x H = E_m\,\Ga_{nm},
\label{eq:manes-de-Triton} 
\end{equation}
form a doubly indexed orthogonal basis for the space of functions on
phase space. These functions describe stationary states when $n = m$;
when $n\ne m$ they describe transitions between pairs of states. Note 
that $\Ga_{nm} = \overline{\Ga_{mn}}$ by~\eqref{eq:rollo-habitual}.
\item
The sequence of those eigenvalues gives precisely the spectrum of~$H$.
\item
With the normalization
$$
\int |\Ga_{nm}|^2 \,\frac{dq\,dp}{(2\pi)^{3N}} = 1,
$$
we obtain
\begin{equation}
\Ga_{nm} \x \Ga_{kl} = \Ga_{nl}\,\dl_{mk} \word{and the identity}
\int \Ga_{mn}(q;p)\, \frac{dq\,dp}{(2\pi)^{3N}} = \dl_{mn},
\label{eq:animal-chico} 
\end{equation}
consistently with \eqref{eq:rollo-habitual}
and~\eqref{eq:manes-de-Triton}. Clearly $\Ga_{nn}/(2\pi)^{3N}$ is the
Wigner quasiprobability corresponding to the state of energy~$E_n$.
\item
Also, the following closure relations hold:
$$
\sum_n \Ga_{nn}(q;p) = 1; \qquad
\sum_{m,n} \Ga_{nm}(q;p) \Ga_{mn}(q';p') 
= (2\pi)^{3N} \,\dl(q - q') \,\dl(p - p').
$$
\item
Finally, we can write
$$
H = \sum_n E_n\,\Ga_{nn}.
$$
\end{itemize}
\end{thm}

\begin{cor} 
\label{cr:minimum-effort}
For any normalized state $P_N$ and any Hamiltonian $H$, if $E_0$
denotes the energy of the ground state corresponding to~$H$, then:
$$
E_0 \leq \int H(q;p) P_N(q;p) \,dq\,dp.
$$
Equality is reached if and only if $P_N = P_\gs$, the Wigner
distribution for the ground state.
\end{cor} 

The validity of this variational principle is of course not restricted
to electronic systems.

\begin{proof}
Any function~$F$ of the phase space variables may be expanded in a
double series of eigentransitions corresponding to~$H$. In particular,
\begin{equation}
P_N = \sum_{m,n} c_{mn} \Ga_{mn},
\label{eq:animal-grande} 
\end{equation}
where:
$$
c_{mn} = \int P_N(q;p) \Ga_{nm}(q;p) \,\frac{dq\,dp}{(2\pi)^{3N}}.
$$
Thus
$$
\int H(q;p) P_N(q;p) \,\frac{dq\,dp}{(2\pi)^{3N}}
= \int (H \x P_N)(q;p) \,\frac{dq\,dp}{(2\pi)^{3N}}
= \sum_n E_n\,c_{nn},
$$
where we have used \eqref{eq:rollo-habitual}, \eqref{eq:animal-chico}
and~\eqref{eq:animal-grande}. Now take for $F$ the symbol $W_N$ of a
pure state, so that $W_N = W_N \x W_N$. An immediate calculation
gives $c_{nn} = \sum_m |c_{mn}|^2 \geq 0$. The rest is obvious.
\end{proof}

\section{The main theorems}
\label{sec:big-picture}

Let us invoke the classical Hamiltonian of an electronic system under
the form:
\begin{equation}
H = T + V_\ee + V_\ext,
\label{eq:nonrel-description} 
\end{equation}
where
$$
T = \sum_{i=1}^N \frac{|\p_i|^2}{2}; \qquad
V_\ee = \sum_{i<j} \frac{1}{|\q_i - \q_j|}; \qquad
V_\ext = \sum_{i=1}^N V(\q_i;\p_i).
$$
Here $V_\ext$ denotes the external ``potential'' (e.g., due to the
nucleus in an atom plus an external magnetic field). It is clear that
the energy of a $N$-electronic system is a functional of~$d_2$. But we
can prove more.

\begin{thm} 
The many-body ground state of the system is determined by its 1-body
quasidensity.
\end{thm}

\begin{proof}
Like in DFT, we argue by contradiction. Suppose that we have two
different (by more than a constant) external potentials $V_\ext$,
$V'_\ext$ acting on our electronic system, with corresponding ground
Wigner states $P_\gs$, $P'_\gs$ (assumed different) and respective
ground state energies $E_0$, $E'_0$, whose associated 1-body
quasidensity is the same. Then, by the variational principle on phase
space of the previous section, we get
\begin{align*}
E'_0 &= \int H' P'_\gs \,dq\,dp < \int H' P_\gs \,dq\,dp
\\
&= \int (H + V'_\ext - V_\ext) P_\gs \,dq\,dp
= E_0 + \int (V'_\ext - V_\ext) P_\gs \,dq\,dp.
\end{align*}
But
$$
\int (V'_\ext - V_\ext) P_\gs \,dq\,dp
= \int (V'_\ext - V_\ext)(\q;\p)\, d_\gs(\q;\p) \,d\q\,d\p.
$$
The same argument shows that
$$
E_0 < E'_0 + \int (V_\ext - V'_\ext)(\q;\p)\,d_\gs(\q;\p) \,d\q\,d\p.
$$
and we obtain a contradiction. Thus $d_\gs$ fixes $P_\gs$ (and thereby
the expected value of any observable in the ground state).
\end{proof}

Next we avoid $V$-representability problems by a constrained-search
definition of the energy functional.

\begin{thm} 
There exists a functional $\A$ of the electronic quasiprobability~$d$,
such that:
$$
E_0 \leq \int V(\q;\p) d(\q;\p) \,d\q\,d\p
+ \int \frac{|\p|^2}{2}\, d(\q;\p) \,d\q\,d\p + \A[d] =: \E[d].
$$
Moreover, if $d_\gs$ is the quasidensity corresponding to the ground
state, then:
$$
E_0 = \int V(\q,\p) d_\gs(\q,\p) \,d\q\,d\p
+ \int \frac{|\p|^2}{2}\, d_\gs(\q,\p) \,d\q\,d\p + \A(d_\gs)
= \E(d_\gs).
$$
\end{thm}

\begin{proof}
Let $\A[d]$ be
\begin{equation}
\min \int P_d(q;p) V_\ee(q;p) \,dq\,dp,
\label{eq:make-or-break} 
\end{equation}
where $P_d$ runs through every Wigner $N$-distribution (representing
mixed states in general) giving the fixed quasidensity~$d$. Let
$P_d^\mini$ be the one attaining this minimum. (Such $P_d$ form a
compact convex set, with respect to a topology that makes 
$P_d \mapsto \int P_d V_\ee$ a continuous linear form, so its minimum
will be attained.) The variational principle says that
\begin{equation}
\int V(\q;\p) d(\q;\p) \,d\q\,d\p
+ \int \frac{|\p|^2}{2} \pi(\p) \,d\p + \A[d]
= \int H(q;p) P_d^\mini(q;p) \,dq\,dp \geq E_0.
\label{eq:hora-de-la-verdad} 
\end{equation}
In particular,
$$
E_0 = \int H(q;p) P_\gs(q;p) \,dq\,dp
\leq \int H(q;p) P_{d_\gs}^\mini(q;p) \,dq\,dp.
$$
This gives
$$
\int P_\gs(q;p) V_\ee(q;p) \,dq\,dp
\leq \int P_{d_\gs}^\mini(q;p) V_\ee(q;p) \,dq\,dp.
$$
On the other hand, by definition, the reverse inequality holds:
$$
\int P_{d_\gs}^\mini(q;p) V_\ee(q;p) \,dq\,dp
\leq \int P_\gs(q;p) V_\ee(q;p) \,dq\,dp.
\eqno \qed
$$
\hideqed
\end{proof}

The minimization has been carried out in two stages. First we perform
a search constrained by the trial quasidensity~$d$. In the second
step, expression~\eqref{eq:hora-de-la-verdad} is minimized. By
Corollary~\ref{cr:minimum-effort}, $P_\gs = P_{d_\gs}^\mini$, which
means that $P_\gs$ can be obtained from $P^\mini_{d_\gs}$ directly
even if the external potential is unknown. As in Levy's formulation of
the Hohenberg--Kohn functional~\cite{Levy82}, $\A$ is universal, i.e.,
independent of~$V$; nor is a ground state $V$-representability
condition required. Thus our variational principle escapes the major
problems of the original DFT one -- see the discussion
in~\cite[Chap.~33]{BlanchardB03}. Systems with external potentials
depending on momenta (as with orbital magnetism) fall into its purvey.
Also, in the variation above we need not restrict to one-body Wigner
quasiprobabilities corresponding to pure states; only finiteness
conditions, such as that $d$ should belong to the domain of the
kinetic energy,
\begin{equation}
T[d] = \int \frac{|\p|^2}{2}\, d(\q;\p) \,d\q\,d\p < \infty,
\label{eq:memento} 
\end{equation}
are implicit. Thus our apparent restriction to nondegenerate ground
states is merely an inessential notational simplification: WDFT is an
ensemble functional theory able to deal with degenerate ground states.
For the same reason, $\A[d]$ is a convex functional, and therefore
$\E[d]$ too is convex.

\subsection{Exact requisites for the quasidensity functional}
\label{ssc:disturbare}

In this subsection we collect a number of properties and conditions
on $\A[d]$ within exact WDFT.

\begin{itemize}
\item
A Legendre transform variant of $\A$ exists:
$$
\A_L[d] = \sup \Bigl[ E[V_\ext] - \int
\bigl(\half|\p|^2 + V_\ext(\q;\p)\bigr) d(\q;\p) \,d\q\,d\p\, \Bigr].
$$
The supremum is taken over all possible choices of the external
potential. The Legendre transform allows a reformulation of the 
previous variational theorems, in the spirit of~\cite{AyersGL06}.

\item
Let $V_\ext$ depend on a parameter~$\la$. Then we denote the
Hamiltonian by~$H_\la$. Consider the associated eigenvalue problem:
$$
H_\la \x \Ga_\la = E_\la\,\Ga_\la = \Ga_\la \x H_\la,
$$
with $\Ga_\la$ being a normalized Wigner distribution. The assertion 
that
\begin{equation}
\frac{dE_\la}{d\la} = \int \pd{H_\la(q;p)}{\la}\,\Ga_\la(q;p) \,dq\,dp
\label{eq:Hellmann-Feynman} 
\end{equation}
is the Hellmann--Feynman theorem in phase space quantum mechanics.

\begin{proof}
Indeed, $E_\la = \int H_\la(q;p) \,\Ga_\la(q;p) \,dq\,dp$ implies
$$
\frac{dE_\la}{d\la} = \int \pd{H_\la(q;p)}{\la}\,\Ga_\la(q;p) \,dq\,dp
+ \int H_\la(q;p) \,\pd{\Ga_\la(q;p)}{\la} \,dq\,dp.
$$
However,
$$
\int \! H_\la(q;p) \,\pd{\Ga_\la(q;p)}{\la} \,dq\,dp
= \int \! H_\la(q;p) \x \pd{\Ga_\la(q;p)}{\la} \,dq\,dp
= E_\la \int \pd{\Ga_\la(q;p)}{\la} \,dq\,dp = 0.
$$
The theorem follows.
\end{proof}

\item
Now, let us write
$$
V_{\ext,\la}(\q;\p)
= V_\ext(\q;\p) + \la\r_1\.\nabla_{\!\q} V_\ext(\q;\p)
+ \la\r_2\.\nabla_{\!\p} V_\ext(\q;\p) +\cdots
$$
for arbitrary vectors $\r_1,\r_2$. Thus,
by~\eqref{eq:Hellmann-Feynman} and homogeneity of space,
$$
\int d(\q;\p)(\nabla_{\!\q} + \nabla_{\!\p})V_\ext(\q;\p) \,d\q\,d\p
= 0.
$$
Similarly, by isotropy of space,
$$
\int d(\q;\p)(\q\x\nabla_{\!\q} + \p\x\nabla_{\!\p}) V_\ext(\q;\p)
\,d\q\,d\p = 0.
$$
Only stationarity of the state is required. These results follow as
well from the minimum principles of the previous section.

\item
For stationary states,
\begin{align*}
E &= - \int \frac{|\p|^2}{2}\, d(\q;\p) \,d\q\,d\p
\\
&= \frac{1}{2} \biggl( \int V(\q) d(\q;\p) \,d\q\,d\p
+ \int \frac{d_2(\q_1,\q_2;\p_1,\p_2)}{|\q_1 - \q_2|}
\,d\q_1\,d\q_2\,d\p_1\,d\p_2 \biggr).
\end{align*}
This is the virial theorem in phase space quantum mechanics; we
consider here pure Cou\-lomb systems. As a corollary, we get for the
ground state:
$$
E_0 = -\int \frac{|\p|^2}{2}\, d_\gs(\q,\p) \,d\q\,d\p
= \frac{1}{2} \biggl( \int V(\q,\p) d_\gs(\q,\p) \,d\q\,d\p
+ \A(d_\gs) \biggr).
$$

\item
We should not overlook that the minimization process takes place under
the constraint $\int d(\q;\p) \,d\q\,d\p = N$, which can be
implemented by a Lagrange multiplier. Then one may minimize
$$
\E[d] - \mu \Bigl( N - \int d(\q;\p) \,d\q\,d\p \Bigr).
$$
The multiplier $\mu$ (with a minus sign) is an important physical
parameter, called (Mulliken's) electronegativity. Recall that for a
neutral atom $E_0(N-1) - E_0(N)$ is the ionization potential, and
$E_0(N) - E_0(N+1)$ is the electron affinity. Their average
constitutes a finite-difference approximation to the
electronegativity.%
\footnote{The behaviour of these quantities is of current theoretical
interest as a marker of the limitations of approximate functionals;
see~\cite{CohenMSY08} and references therein. We cannot go into that
here, however.}

\item
Finally, we look at \textit{scaling}. Matters are pretty satisfactory
with~$d$ in this respect. Let $\la > 0$ be a scale factor. We scale
the Wigner distribution by defining 
$P_\la(q;p) := P_\la(\la q;\la^{-1}p)$, another Wigner distribution,
whose scaled quasidensity is
$d_\la(\q;\p) = d(\la\q;\la^{-1}\p)$, yielding the scaled density
$\rho_\la(\q) = \la^3 \rho(\la\q)$. One can show that $d_\la$
represents a Wigner eigenstate of a Hamiltonian of the form
$T + \la V_\ee + \sum_{i=1}^N \la^2 V_d(\q_i;\p_i)$. Now $T[d_\la]$
is $\la^2 T[d]$. As a consequence,
\begin{equation}
\A[d_\la] = \la\A[d];  \word{and trivially}
\pd{\A[d_\la]}{\la} \biggr|_{\la=1} = \A[d].
\label{eq:mas-sabe-el-diablo} 
\end{equation}
The situation in the standard DFT approach with respect to scaling is
much more involved. Denote $Q[\rho] = T[\rho] + V_\ee[\rho]$, the
universal Hohenberg--Kohn--Levy functional. The naive expectations
$T[\rho_\la] = \la^2 T[\rho]$ and $V_\ee[\rho_\la] = \la V_\ee[\rho]$
are both false: one can show that $T[\rho_\la] > \la^2 T[\rho]$, 
$V_\ee[\rho_\la] < \la V_\ee[\rho]$ for $\la < 1$; whereas
$T[\rho_\la] < \la^2 T[\rho]$, $V_\ee[\rho_\la] > \la V_\ee[\rho]$ for
$\la > 1$. Nor is it possible to partition $Q$ into two functionals
in some other way with the desired behaviour~\cite{LevyP85}.

We emphasize that these constraints on~$\A[d]$ are valid for
\textit{arbitrary} quasidensities; this is why they are potentially
useful. Explanation of the good behaviour in this regard of WDFT with
respect to ordinary DFT lies obviously in the exactness of the kinetic
energy functional in the former; whereas in the latter $T[\rho]$ is a
big unknown complicated functional, which ``pollutes'' the Coulomb
energy.
\end{itemize}

In summary, WDFT splits the problems of density functional theory into
the (solved) characterization problem for 1-body Wigner functions and
the determination problem for $\A[d]$, a functional both smaller in
magnitude and less slippery in principle than that of Hohenberg--Kohn
theory; exactness and simplicity of the kinetic energy functional
commend our method. But one needs familiarity with the lore of Wigner
quasiprobabilities.

\section{Getting used to Wigner quasiprobabilities}
\label{sec:settling-in}

\subsection{Harmonium via the Wigner function}
\label{ssc:harmonica}

In order to win intuition on the workings of $\A[d]$, it is good to
study the WDFT functional in an analytically solvable problem. So we
consider two fermions trapped in a harmonic potential well, which
moreover couple to each other with a repulsive Hooke law force; this
is the so-called \textit{harmonium}, or Moshinsky
atom~\cite{Moshinsky68}. The one-dimensional case has been treated on
phase space in~\cite{Dahl09}. Introduce extracule and intracule
coordinates, respectively given by
$$
\vec R = (\q_1 + \q_2)/2, \qquad  \vec\rr = \q_1 - \q_2,
$$
with conjugate momenta
$$
\vec P = \p_1 + \p_2,  \qquad  \vec\pp = (\p_1 - \p_2)/2.
$$
The classical Hamiltonian is given by
$$
H = H_R + H_\rr
= \frac{P^2}{4} + \om^2 R^2 + \pp^2 + (\om^2 - k) \frac{\rr^2}{4};
$$
the last term includes the electronic repulsion $-k\rr^2/4$ (we assume
$0 \leq k < \om^2$; obviously for $k \geq \om^2$ the repulsion between
the particles is so strong that they cannot both remain in the well).
The energy of the ground state is clearly given by
$E_0 = \frac{3}{2}(\om + \sqrt{\om^2 - k}\,)$.

The corresponding Wigner function for the ground state factorizes into
an extracule and an intracule phase space quasidensity:
\begin{align}
& P_\gs(\q_1,\q_2;\p_1,\p_2;\vs_1,\vs_2;\vs_{1'}\vs_{2'})
\notag \\
&\qquad = \frac{1}{2} (\up_1\dn_2 - \dn_1\up_2)
(\up_{1'}\dn_{2'} - \dn_{1'}\up_{2'}) \, \frac{1}{\pi^6}
\exp\Bigl( -\frac{2H_R}{\om} \Bigr)
\exp\Bigl( -\frac{2H_\rr}{\sqrt{\om^2 - k}} \Bigr).
\label{eq:agallas} 
\end{align}
Note the correct normalization
$$
\int P_\gs(\q_1,\q_2;\p_1,\p_2;\vs_1,\vs_2;\vs_1,\vs_2) \,d1\,d2 = 1.
$$

The electron interaction energy can be obtained from the intracule
in~\eqref{eq:agallas}:
$$
-\frac{k}{4\pi^3} \int e^{-2H_\rr/\sqrt{\om^2 - k}}\, \rr^2
\,d\vec\rr \,d\vec\pp = -\frac{3k}{4\sqrt{\om^2 - k}} \,.
$$
We have generalized the \textit{Wigner intracule}, in the terminology
of~\cite{GillCONB06}, there valid only for non-interacting fermions in
the harmonic well. In their notation, after integration over the
angles it is given by
\begin{equation}
W(u,v) = \frac{2}{\pi} u^2 v^2\, e^{-\sqrt{\om^2 - k}\,u^2/2}
e^{-v^2/2\sqrt{\om^2 - k}}.
\label{eq:arma-virumque} 
\end{equation}
Here $v = 2\pp$. By the way, the above type of calculation applies
quite generally, not only for Hartree--Fock (Slater determinant)
states, as declared in that paper.

Now that we are at that, let us compute the relative-motion and
centre-of-mass components of the kinetic energy~$T_0$ and the
confinement energy. For the former we get, respectively,
$$
\frac{1}{\pi^3} \int e^{-2H_\rr/\sqrt{\om^2 - k}}\,
\pp^2 \,d\vec\pp\,d\vec\rr = \frac{3\sqrt{\om^2 - k}}{4}  \word{and}
\frac{1}{4\pi^3} \int e^{-2H_R/\om}\, P^2 \,d\vec P\,d\vec R
= \frac{3\om}{4},
$$
and the virial theorem is fulfilled, since $E_0 = 2T_0$. For the
latter,
$$
\frac{\om^2}{4\pi^3} \int e^{-2H_\rr/\sqrt{\om^2 - k}}\,
\rr^2 \,d\vec\rr\,d\vec\pp = \frac{3\om^2}{4\sqrt{\om^2 - k}}
\word{and}
\frac{\om^2}{\pi^3} \int e^{-2H_R/\om}\, R^2 \,d\vec R\,d\vec P
= \frac{3\om}{4}.
$$
In all,
$$
\frac{3\om}{2} + \frac{3\sqrt{\om^2 - k}}{4}
+ \frac{3\om^2}{4\sqrt{\om^2 - k}} - \frac{3k}{4\sqrt{\om^2 - k}}
= \frac{3}{2} (\om + \sqrt{\om^2 - k}\,),  \word{indeed.}
$$
Since the value of the centre-of-mass energy is preordained, it should
be clear that only (the two marginals of)
expression~\eqref{eq:arma-virumque} is employed.

\subsection{The Wigner 1-quasiprobability}
\label{ssc:Gilli}

Also from~\eqref{eq:agallas}, one gets
\begin{align}
& P_\gs(\q_1,\q_2;\p_1,\p_2) = \frac{1}{\pi^6} \exp\Bigl( 
-\frac{1}{2\om} (p_1^2 + p_2^2 + 2\p_1\.\p_2)
- \frac{\om}{2} (q_1^2 + q_2^2 + 2\q_1\.\q_2) \Bigr)
\notag \\
&\qquad \x \exp\Bigl(
-\frac{1}{2\sqrt{\om^2 - k}} (p_1^2 + p_2^2 - 2\p_1\.\p_2)
- \frac{\sqrt{\om^2 - k}}{2} (q_1^2 + q_2^2 - 2\q_1\.\q_2) \Bigr)
\label{eq:velis-quod-potis} 
\end{align}
Now we compute the 1-body phase space quasidensity for the ground
state. One obtains:
\begin{align}
& d_\gs(\q;\p;\vs;\vs') 
\notag \\
&\qquad = \frac{\up\up' + \dn\dn'}{2} \, \frac{2}{\pi^3} \biggl( 
\frac{4\om\sqrt{\om^2 - k}}{(\om + \sqrt{\om^2 - k})^2} \biggr)^{3/2}
e^{-2q^2\om\sqrt{\om^2 - k}/(\om + \sqrt{\om^2 - k}\,)}
e^{-2p^2/(\om + \sqrt{\om^2 - k})},
\label{eq:ubi-Roma} 
\end{align}
with marginal distributions
\begin{align}
\rho_\gs(\q) &= 2 \biggl(\frac{\om}{\pi}\biggr)^{3/2}
\biggl( \frac{2\sqrt{\om^2 - k}}{\om + \sqrt{\om^2 - k}} \biggr)^{3/2}
e^{-2q^2\om\sqrt{\om^2 - k}/(\om + \sqrt{\om^2 - k})},
\label{eq:por-viejo} 
\\ 
\pi_\gs(\p) &= 2(\om\pi)^{-3/2}
\biggl( \frac{2\om}{\om + \sqrt{\om^2 - k}} \biggr)^{3/2}
e^{-2p^2/(\om +\sqrt{\om^2 - k})}.
\notag
\end{align}
The normalization is now
$$
\int d_\gs(\q;\p;\vs;\vs) \,d\q\,d\p\,d\vs
= \int \rho_\gs(\q) \,d\q = \int \pi_\gs(\p) \,d\p = 2.
$$
{}From the above we can easily recompute the kinetic and confinement
energy parts:
$$
T_0 = \frac{1}{2} \int |\p|^2 \pi_\gs(\p) \,d\p
= 3\,\frac{\om + \sqrt{\om^2 - k}}{4}; \quad 
V_{\ext,0} = \frac{\om^2}{2} \int |\q|^2 \rho_\gs(\q) \,d\q
= 3\,\frac{\om^2 + \om\sqrt{\om^2 - k}}{4\sqrt{\om^2 - k}} \,.
$$
This yields $V_{\ee,0} = -3k/4\sqrt{\om^2 - k}$, as we had obtained
directly.

Note that $d_\gs(0;0) < 2\pi^{-3}$ for $k > 0$; there is a telltale
\textit{tassement} of the Wigner quasiprobability with respect to what
is typical for ground pure 1-particle states. Note also that
\begin{align*}
\int d_\gs(\q;\p)^2 \,d\q\,d\p &= \frac{4}{\pi^6} \biggl(
\frac{4\om\sqrt{\om^2 - k}}{(\om + \sqrt{\om^2 - k})^2} \biggr)^3
\! \int e^{-4q^2\om\sqrt{\om^2 - k}/(\om + \sqrt{\om^2 - k}\,)}
e^{-4p^2/(\om + \sqrt{\om^2 - k})} \,d\q\,d\p
\\
&= \frac{4}{(2\pi)^3} \biggl( 
\frac{4\om\sqrt{\om^2 - k}}{(\om + \sqrt{\om^2 - k})^2} \biggr)^{3/2}
\leq \frac{4}{(2\pi)^3}.
\end{align*}
This is in agreement with~\eqref{eq:nesciunt-qui-faciunt}. It is clear
that this $d_\gs$ cannot correspond to a Hartree--Fock (HF) state
unless $k = 0$. 

We remark that the exact one-particle density matrix of
\cite[Eq.~(2--68)]{Davidson76} and~\cite{AmovilliM03} for this problem
is obtained from $d_\gs$ simply by using the inversion
formula~\eqref{eq:pasodoble}.

\subsection{Pairs density}
\label{ssc:Gillipar}

Note that
$$
\rho(\q_1)\rho(\q_2) = 4\bigl( 
2\om\sqrt{\om^2 - k} \big/ \pi(\om + \sqrt{\om^2 - k}\,) \bigr)^3
e^{-\frac{2\om\sqrt{\om^2 - k}}{\om + \sqrt{\om^2 - k}}
(q_1^2 + q_2^2)}.
$$
On the other hand, by integrating out the momenta
in~\eqref{eq:velis-quod-potis},
\begin{equation}
\rho_2(\q_1,\q_2) 
= \biggl( \frac{\om\sqrt{\om^2-k}}{\pi^2} \biggr)^{3/2}
e^{-\shalf(\om + \sqrt{\om^2-k})(q_1^2 + q_2^2)}
e^{-(\om - \sqrt{\om^2-k})\,\q_1\.\q_2}.
\label{eq:cano} 
\end{equation}
With spin components:
$$
\rho_2(1,2) = \frac{1}{2} (\up_1\dn_2 - \dn_1\up_2)
(\up_1\dn_2 - \dn_1\up_2) \,\rho_2(\q_1,\q_2).
$$
Finally, we obtain
$$
\frac{4\,\rho_2(\q_1,\q_2)}{\rho(\q_1)\rho(\q_2)}
= \biggl( \frac{(\om + \sqrt{\om^2 - k}\,)^2}{4\om\sqrt{\om^2 - k}}
\biggr)^{3/2} e^{-(\om - \sqrt{\om^2 - k}\,)\,\q_1\.\q_2}
e^{-\frac{(\om - \sqrt{\om^2 - k}\,)^2}{2(\om + \sqrt{\om^2 - k})}
(q_1^2 + q_2^2)}
$$
without recourse to wavefunctions, density matrices, or the like. As
was pointed out in~\cite{Davidson76}, besides the angular correlation
in the pair distribution, favouring $\q_1 = -\q_2$, which was to be
expected, we see a \textit{contraction} relative to the uncorrelated
distribution.

The above calculation can be organized in a better way, by integrating
in~\eqref{eq:agallas}
\begin{align*}
\rho_2(\vec R,\vec\rr) &= \frac{1}{\pi^6} \int 
\exp\biggl( -\frac{2H_R}{\om} \biggr)
\exp\biggl( -\frac{2H_\rr}{\sqrt{\om^2 - k}} \biggr)
\,d\vec P\,d\vec\pp
\\
&= \biggl( \frac{2\om}{\pi} \biggr)^{3/2} e^{-2\om R^2}
\biggl( \frac{\sqrt{\om^2 - k}}{2\pi} \biggr)^{3/2}
e^{-\sqrt{\om^2 - k}\,\rr^2/2} =: E(\vec R)\,I(\vec\rr),
\end{align*}
coincident with~\eqref{eq:cano}.

\subsection{On the functional theory}
\label{ssc:harmful-functional}

Now, according to the tenets of functional theories, $V_\ext$ must be
a functional of~$\rho_\gs$ or~$\pi_\gs$; depending on whether we
employ ordinary DFT or Henderson's variant based on the momentum
density~\cite{Henderson81}. Dahl argues in~\cite{Dahl09} that (modulo
an arbitrary constant $V_0$) one can indeed recover the potential,
granted the harmonic form for it. But in general one would only be
able to estimate the second derivative of the confining potential at
the origin. So the Kohn--Hohenberg method remains nonconstructive,
even for the harmonic interaction between the fermions.

What does this example teach us about $\A[d]$ when $V_\ee$ is of the
harmonic form? One could be seduced by the following chain of
reasoning. Note first that
\begin{align*}
\rho_0 := \rho_\gs(0) &= 2\biggl( \frac{\om}{\pi} \biggr)^{3/2}
\biggl(\frac{2\sqrt{\om^2 - k}}{\om + \sqrt{\om^2 - k}} \biggr)^{3/2};
\\
\pi_0 := \pi_\gs(0) &= 2\biggl( \frac{1}{\om\pi} \biggr)^{3/2}
\biggl( \frac{2\om}{\om + \sqrt{\om^2 - k}} \biggr)^{3/2};
\end{align*}
from which 
\begin{equation}
\om = \om(k;\rho_0/\pi_0)
\label{eq:ubi-alium} 
\end{equation}
is obtained by solving a simple algebraic equation. Therefore for
such~$\om$ we have
$$
V(\q) = \frac{\om(\om + \sqrt{\om^2 - k}\,)}{4\sqrt{\om^2 - k}}
\log \frac{\rho_0}{\rho_\gs(\q)} \,;  \qquad  \frac{p^2}{2} 
= \frac{\om + \sqrt{\om^2 - k}}{4} \log \frac{\pi_0}{\pi_\gs(\p)},
$$
and one could surmise that
\begin{align*}
\A[d] &= E_0 - \int V(\q) \rho(\q) \,d\q
- \int \frac{p^2}{2} \pi(\p) \,d\p
\\[\jot]
&= \frac{\om + \sqrt{\om^2 - k}}{4} \biggl[ 6
+ \int \pi(\p) \log \frac{\pi(\p)}{\pi_0} \,d\p
+ \frac{\om}{\sqrt{\om^2 - k}} \int 
\rho(\q) \log \frac{\rho(\q)}{\rho_0} \,d\q \biggr],
\end{align*}
where in the last line all reference to the external potential has
been banished. Even so, we have not obtained the universal Wigner
functional for harmonic interparticle actions, because we have unduly
restricted the variation defining~$\A[d]$. In spite of this, for
confined two-particle systems the above formula doubtless constitutes
a good approximation -- in parallel to what was shown
in~\cite{Schindlmayr99} in the context of ordinary DFT.

On a more narrow definition, restricting ourselves to harmonium, and
given that we know the functional forms of the Wigner one-body
quasidensity and the Wigner intracule, we certainly can determine the
strength of the particle-particle interaction from $d$ -- it is enough
to look at $d_\gs(0;0)$ -- and thus the interaction energy.

\section{A Gaussian interlude}
\label{sec:ludi-cereales}

Beyond the ease of calculations with Gaussians quasiprobabilities,
there are pertinent reasons, from the use of Gaussian basis sets in
standard quantum chemistry~\cite{SzaboO89,Jensen07}, and from
entanglement theory~\cite{DahlMWS06}, that make it imperative to learn
how to manipulate them. Also, phase space functions with negative
regions can be reached by means of transitions between Gaussians. So
we need to characterize those transitions. This is taken up in this
section.

We consider here $s$-type Gaussians centered at the origin; any others
can be obtained by derivation -- see the appendix
in~\cite{SpringborgD87} -- and translation. Not every Gaussian on
phase space can represent a quantum state. Suppose that, with
$u \equiv (q,p)$, we do have a Gaussian
\begin{equation}
P_F(u) = \pi^{-n}(\det F)^{\half}\exp(-u^tFu), \word{thus
normalized by} \int P_F(u)\,d^{2n}u = 1;
\label{eq:quasi-Gauss} 
\end{equation}
and suppose we want it to represent a \textit{pure} state. Then $F$,
beyond being positive definite, must be
\textit{symplectic}~\cite{Littlejohn86}. This will entail 
$\det F = 1$. Recall that, if $J$ denotes the canonical complex
structure,
$$
J := \twobytwo{0_n}{1_n}{-1_n}{0_n},
$$
the matrix $F$ is symplectic if $FJF^t = J$. Any positive definite
symplectic matrix can be factorized as $F = S^t S$ where $S$ and~$S^t$
are symplectic, too. In particular, such matrices $F$ are
symplectically congruent to the identity. The space of such Gaussians
is of dimension $n^2 + n$. In our case, $n = 3N$.

A Gaussian on phase space represents a \textit{mixed} state if $F$ is
symplectically congruent to
$$
\diag(\la_1,\dots,\la_n,\la_1,\dots,\la_n)
$$
with $0 < \la_i < 1$. This was found in~\cite{Titania}. The space of
such Gaussians is of dimension $2n^2 + n$.

\smallskip

Given a symplectic (and positive definite) $F$, we can partition it 
into four $n \x n$ blocks:
$$
F = \twobytwo{A}{B}{C}{D^{-1}}.
$$
Here the diagonal blocks are invertible, which is not always the case
for general symplectic matrices; we choose the notation $D^{-1}$ for
convenience. Moreover $A = A^t$, $D = D^t$, $C = B^t$. Now,
$$
\twobytwo{A}{B}{B^t}{D^{-1}} J \twobytwo{A}{B}{B^t}{D^{-1}} = J
\word{implies} BD = DB^t,\ AB^t = BA, \ A = D + B D B^t.
$$
We can know which wavefunction (also of Gaussian form) a Gaussian pure
state comes from: $P_F$ is the Wigner function corresponding to
$\ketbra{\Psi}{\Psi}$ where, up to a constant phase factor,
$$
\Psi(q) = \frac{(\det D)^{1/4}}{\pi^{3N/4}}
\exp\bigl(-\half q\. D q - \tihalf q\. BD q \,\bigr).
$$
In view of the above, $D$ and $BD$ are respectively positive and
symmetric. For instance, the ground state wavefunction of the
two-electron system considered in the previous subsection is given by
\begin{equation}
\Psi_\gs(\vec R,\vec\rr\,)
= \frac{\up_1\dn_2 - \dn_1\up_2}{\sqrt{2}}\,
\biggl( \frac{\om}{\pi} \biggr)^{3/4} e^{-\om R^2}
\biggl( \frac{\sqrt{\om^2 - k}}{\pi} \biggr)^{3/4}
e^{-\squarter\sqrt{\om^2 - k}\,\rr^2} \,.
\label{eq:dia-sin-agua} 
\end{equation}
A similar formula holds in momentum space. 

\vspace{6pt}

It is of high interest to find the \textit{Wigner transition}
corresponding to $\ketbra{\Psi_1}{\Psi_2}$, where $\Psi_1$ and
$\Psi_2$ are Gaussians; they are bound to play an important role in
calculations with Gaussian orbitals in quantum chemistry. All the
information about $\Psi_1$, $\Psi_2$ is contained in their phase space
partners $P^1 \equiv P_{F_1}$, $P^2 \equiv P_{F_2}$, so we can
characterize the transitions from the parameters of the quadratic
symplectic forms $F_1$, $F_2$.

\begin{thm} 
The Wigner transition $P^{12}(q;p) \equiv P^{12}(u)$ between two
pure-state Gaussian quasidensities $P^1(u)$ and $P^2(u)$ given by
\eqref{eq:quasi-Gauss} is of the form
$$
P^{12}(u) = \frac{(\det D_1)^{1/4}\,(\det D_2)^{1/4}}
{\pi^{3N}\,(\det D_{12})^{1/2}}\, e^{- u\. F_{12} u},
\word{where}
F_{12} := \twobytwo{A_{12}}{B_{12}}{B_{12}^t}{D_{12}^{-1}}
$$
is a complex symmetric and \emph{symplectic} matrix with positive
definite real part, whose components are given by
\begin{align}
D_{12} &:= \half \bigl( D_1 + D_2 + i(B_1D_1 - B_2D_2) \bigr),
\nonumber \\
B_{12} &:= \bigl( -\tihalf(D_1 - D_2) + \half(B_1D_1 + B_2D_2) \bigr)
D_{12}^{-1},
\nonumber \\
A_{12} &:= D_{12} + B_{12} D_{12} B_{12}^t \,.
\label{eq:bobs-your-uncle} 
\end{align}
\end{thm}

\begin{proof}
The integral
$$
\braket{\Psi_2}{\Psi_1}
= \frac{(\det D_1)^{1/4}\,(\det D_2)^{1/4}}{\pi^{3N/2}}
\int e^{-\shalf q\.(D_1 + D_2 + i(B_1D_1 - B_2D_2))q} \,d^{3N}q
$$
converges absolutely, since the complex symmetric matrix 
$D_{12} = \half(D_1 + D_2 + i(B_1D_1 - B_2D_2))$ has positive definite
real part $\half(D_1 + D_2)$, and in particular $D_{12}$ is
invertible. Note that
$B_{12}D_{12} = -\tihalf \bigl( D_1 - D_2 + i(B_1D_1 + B_2D_2) \bigr)$
is also symmetric. 

The Wigner transition is $P^{12} = P_{\Psi_1\Psi_2}$, given
by~\eqref{eq:stoking-more-trouble}. Explicitly,
\begin{align*}
P^{12}(q;p) &= \frac{(\det D_1)^{1/4}\,(\det D_2)^{1/4}}{\pi^{9N/2}}
\\
&\qquad \x \int e^{-\shalf(q - z)\.(D_1 + iB_1D_1)(q - z)}
e^{-\shalf(q + z)\.(D_2 - iB_2D_2)(q + z)} e^{2i(p\.z)} \,d^{3N}z
\\
&= \frac{(\det D_1)^{1/4}\,(\det D_2)^{1/4}}{\pi^{9N/2}}\, 
e^{- q\. D_{12} q}
\int e^{- z\. D_{12} z} e^{2iz\.(p + B_{12}D_{12} q)} \,d^{3N}z
\\
&= \frac{(\det D_1)^{1/4}\,(\det D_2)^{1/4}}
{\pi^{3N}\,(\det D_{12})^{1/2}}\, e^{- q\. D_{12} q}\,
e^{-(p + B_{12}D_{12} q)\.D_{12}^{-1}(p + B_{12}D_{12} q)}.
\end{align*}
Here $(\det D_{12})^{1/2}$ means the branch of the square root of
$\det D_{12}$ that is positive when $D_{12}$ is positive definite. We
have used the standard formula for a Gaussian integral, that is
straightforward when $D_{12}$ is positive definite, and extends by
analytic continuation to the case when the real part
$\half(D_1 + D_2)$ is positive definite \cite[Appendix~A]{Folland89}.

The formulas \eqref{eq:bobs-your-uncle} show that the assembled matrix
$F_{12}$ is indeed (complex) symplectic. The Wigner transition
$P^{12}(u)$ is moreover square-summable, so that $F_{12}$ itself has a
positive definite real part.

Whether or not the obvious reciprocal of this result holds is still an
open question.
\end{proof}

Notice in passing that when $D_1 = D_2 = D$ and $B_1 = B_2 = B$, then
$D_{12} = D$ also, and we of course recover
$$
P(q;p) = \pi^{-3N}\, e^{-q\.Aq} e^{-p\.D^{-1}p} e^{-q\.(B + B^t)p}.
$$
For the needs of quantum chemistry, it is largely enough to consider
\textit{real} wavefunctions of Gaussian form. If $B_1 = B_2 = 0$, then
$D_{12} = \half(D_1 + D_2)$, $B_{12} = -i(D_1 - D_2)(D_1 + D_2)^{-1}$,
and $A_{12} = D_1(D_1 + D_2)^{-1}D_2 + D_2(D_1 + D_2)^{-1}D_1$, so
that
\begin{align*}
P^{12}(q;p) 
&= \frac{2^{3N/2}\,(\det D_1)^{1/4}\,(\det D_2)^{1/4}}
{\pi^{3N}\,(\det(D_1 + D_2))^{1/2}}\,
e^{-q\.(D_1(D_1 + D_2)^{-1}D_2 + D_2(D_1 + D_2)^{-1}D_1)q}
\\
&\qquad \x e^{-2p\.(D_1 + D_2)^{-1}p}\,
e^{2iq\.((D_2 - D_1)(D_1 + D_2)^{-1} + (D_1 + D_2)^{-1}(D_2 - D_1))p}.
\end{align*}
The interesting new thing is the last factor: this will allow Wigner
functions associated to the interference of $P^1$ and~$P^2$ to become
negative at some places. Also, depending on the overlap of $P^1$
with~$P^2$, they will exhibit damped oscillations when both $p$
and~$q$ are large. A simple example with $N = 1$ will soon be useful.
To
\begin{align*}
P^{1,2}(\q;\p) &= \frac{1}{\pi^3} e^{-\a_{1,2}q^2} e^{-p^2/\a_{1,2}}
\word{there corresponds} \Psi_{1,2}(\q)
= \Bigl( \frac{\a_{1,2}}{\pi} \Bigr)^{3/4} e^{-\shalf\a_{1,2}q^2} \,;
\\
\intertext{and then}
P^{12}(\q;\p) &= \pi^{-3} \Bigl( \frac{\a_1\a_2}{\pi^2} \Bigr)^{3/4}
\int e^{-\shalf\a_1(q - z)^2} e^{-\shalf\a_2(q + z)^2} e^{2i\p\.\z}
\,d\z
\\
&= \pi^{-3} \biggl( \frac{4\a_1\a_2}{(\a_1 + \a_2)^2} \biggr)^{3/4}
e^{-2q^2\a_1\a_2/(\a_1 + \a_2)} e^{-2p^2/(\a_1 + a_2)}
e^{2i(\a_2 - \a_1)/(\a_1 +\ a_2)\q\.\p} \,.
\end{align*}

\section{Atomic Wigner functions}
\label{sec:revving-up}

\subsection{Gaussian approximations}
\label{ssc:Megisto}

In Hartree units the Hamiltonian of a hydrogen-like atom is
\begin{equation*}
H = \frac{|\p|^2}{2} - \frac{Z}{|\q|}.
\end{equation*}
The wave function for its ground state is well known:
\begin{equation}
\psi_{1s}(\q) = \frac{Z^{3/2}}{\sqrt{\pi}}\, e^{-Zr};  \qquad
\psi_{1s}(\p) = \frac{2\sqrt2}{\pi} \frac{Z^{5/2}}{(Z^2 + |\p^2|)^2}.
\label{eq:storm-clouds} 
\end{equation}
Therefore $P_{1s}(\q;\p)$ is given by
\begin{equation}
\frac{Z^3}{\pi^4} \int d\z\,
e^{-Z|\q - \z|} e^{-Z|\q + \z|} e^{-2i\p\.\z} 
= \frac{8Z^5}{\pi^5} \int d\z\, \frac{1}{|\p - \z|^2 + Z^2}
\,\frac{1}{|\p + \z|^2 + Z^2}\, e^{2i\q\.\z}.
\label{eq:sombrios-presagios} 
\end{equation}
For a long time it had not been known how to compute these integrals
in a closed analytical form, although in one dimension an analogous
problem was solved~\cite{Dahl95}; the geometrical treatment via the
Kustaanheimo--Stiefel transformation in~\cite{HAtom} allows only to
recover partial data for hydrogen-like atoms. A nearly closed
analytical form is now given in~\cite{PraxmeyerMW06}.

However, one can approximate~\eqref{eq:storm-clouds} by a sum of $M$
normalized Gaussians, with real coefficients; thus
$$
P_{1s}^{(M)}(\q;\p) = \sum_{i=1}^M c_i^2 P^{ii}(\q;\p)
+ \sum_{i>j}^M c_i c_j (P^{ij} + P^{ji})(\q;\p).
$$
According to what we have seen in the previous section, it ensues that
\begin{align}
P_{1s}^{(M)}(\q;\p) 
&= \frac{1}{\pi^3} \sum_{i=1}^M c_i^2\, e^{-\a_i r^2 - p^2/\a_i}
\nonumber \\
&\quad + \frac{2}{\pi^3} \sum_{i>j}^M c_i c_j 
\biggl( \frac{4\a_i\a_j}{(\a_i + \a_j)^2} \biggr)^{3/4}
e^{-\tfrac{2\a_i\a_j}{\a_i + \a_j}r^2}
e^{-\tfrac{2}{\a_i + \a_j}p^2}
\cos\biggl( \frac{2(\a_i - \a_j)}{\a_i + \a_j} \q\.\p \biggr).
\label{eq:gato-escaldado} 
\end{align}
It is clear that the (exact or approximate) result only depends on
$r,p,\th$, with $\th$ being the angle between $\q$ and~$\p$. It is
then convenient to take $r,p,\th$, together with three auxiliary
angles, as the phase space variables.

Let us briefly consider the case $M = 1$ first. This amounts to taking
a trial state which is exact for an oscillator. For the energy:
\begin{align}
E(\a,Z) &= \frac{1}{\pi^3} \int \biggl(
\frac{p^2}{2} - \frac{Z}{r} \biggr) e^{-\a r^2 - p^2/\a} \,d\q\,d\p
\notag \\
&= \frac{16}{\pi} \biggl( \int r^2 e^{-\a r^2} \,dr
\int \frac{p^4}{2}\, e^{-p^2/\a} \,dp - Z \int r e^{-\a r^2} \,dr
\int p^2 e^{-p^2/\a} \,dp \biggr)
\notag \\
&= \frac{3\a}{4} - 2Z\sqrt{\frac{\a}{\pi}}.
\label{eq:nox-ruit} 
\end{align}
The minimum is found at $\a_\mathrm{opt} = 16Z^2/9\pi$, so the
``equivalent oscillator'' has frequency $\om = 16Z^2/9\pi$ precisely.
It is equal to $-4Z^2/3\pi$, a pretty good shot at the correct
$-Z^2/2$, given the roughness of the approximation. At the origin
$P_{1s}^{(1)}(\q;\p)$ takes the maximum theoretical value~$1/\pi^3$.

In order to visualize the quasiprobability, one considers the function
$F_{1s}^{(1)}(r;p)$ obtained by integrating over all angles and
multiplying by $r^2p^2$. Its maximum for $Z = 1$ is found at
$$
\bigl(1/\sqrt{\a_\mathrm{opt}}, \sqrt{\a_\mathrm{opt}}\bigr)
= (1.33, 0.75)
$$
in Hartree units, that is $1.33$ times the Bohr radius $a_0$ for the
distance from the origin and $0.75\,a_0^{-1}\hbar$ for the momentum.
Its contour map is given in Fig.~1 of~\cite{DahlS82}. This is a rather
featureless everywhere positive function, that gives a poor idea of
distribution of quasiprobabilities in the $\H$-atom.

Let us now try $M = 2$, allowing for oscillations in the $F_{1s}$
function. In view of~\eqref{eq:gato-escaldado} we get
\begin{align*}
P_{1s}^{(2)}(\q;\p) 
&= \frac{c_1^2}{\pi^3}\, e^{-\a_1 r^2 - p^2/\a_1}
+ \frac{c_2^2}{\pi^3}\, e^{-\a_2 r^2 - p^2/\a_2}
\\
&\qquad + \frac{2c_1c_2}{\pi^3} 
\biggl( \frac{4\a_1\a_2}{(\a_1 + \a_2)^2} \biggr)^{3/4}
e^{-\tfrac{2\a_1\a_2}{\a_1 + \a_2}r^2} e^{-\tfrac{2}{\a_1 + \a_2}p^2}
\cos\biggl( \frac{2(\a_1 - \a_2)}{\a_1 + \a_2} \q\.\p \biggr).
\end{align*}
There is no reason for $c_1$ and $c_2$ to be of the same sign.
However, it is intuitively clear that combinations of the same sign
are energetically preferable; in particular, among the orthogonal
combinations
\begin{align*}
P_{g,u}^{(2)}(\q;\p) 
&= \frac{1}{2 + 2\bigl(\frac{4\a_1\a_2}{(\a_1 + \a_2)^2}\bigr)^{3/4}}
\Biggl[ \frac{1}{\pi^3}\, e^{-\a_1 r^2 - p^2/\a_1}
+ \frac{1}{\pi^3}\, e^{-\a_2 r^2 - p^2/\a_2}
\\
&\qquad \pm \frac{2}{\pi^3} 
\biggl( \frac{4\a_1\a_2}{(\a_1 + \a_2)^2} \biggr)^{3/4}
e^{-\tfrac{2\a_1\a_2} {\a_1 + \a_2}r^2} e^{-\tfrac{2}{\a_1 + \a_2}p^2}
\cos\biggl( \frac{2(\a_1 - \a_2)}{\a_1 + \a_2} \q\.\p \biggr) \Biggr],
\end{align*}
$P_g$ will have the lower energy. We content ourselves with studying
the radial phase space function
\begin{align*}
F_{1s}^{(2)}(r,p) 
&= 8\pi^2 r^2 p^2 \int_0^\pi P_{1s}^{(2)}(r,p,\th) \sin\th \,d\th
= \frac{16}{\pi} r^2 p^2 \bigl(c_1^2 e^{-\a_1 r^2 - p^2/\a_1}
+ c_2^2 e^{-\a_2 r^2 - p^2/\a_2} \bigr)
\\
&\qquad + \frac{32}{\pi} r^2 p^2 c_1c_2 
\biggl( \frac{4\a_1\a_2}{(\a_1 + \a_2)^2} \biggr)^{3/4}
e^{-\tfrac{2\a_1\a_2}{\a_1 + \a_2}r^2} e^{-\tfrac{2}{\a_1 + \a_2}p^2}\,
j_0\biggl( \frac{2(\a_1 - \a_2)}{\a_1 + \a_2} pr \biggr),
\end{align*}
where $j_0(x) = (\sin x)/x$. We expect $P_{1s}^{(2)}(0;0) = \pi^{-3}$,
implying the constraint
$$
1 = c_1^2 + c_2^2 
+ 2c_1c_2\biggl( \frac{4\a_1\a_2}{(\a_1 + \a_2)^2} \biggr)^{3/4}.
$$
Using the formula
$$
\int e^{-\tfrac{2}{\a_1 + \a_2}p^2} 
j_0\biggl( \frac{2(\a_1 - \a_2)}{\a_1 + \a_2} pr \biggr) p^2 \,dp
= \frac{\sqrt{2\pi}(\a_1 + \a_2)^{3/2}}{16}\,
e^{-\tfrac{(\a_1 - \a_2)^2}{2(\a_1 + \a_2)}r^2},
$$
one obtains the radial density of charge:
$$
\rho(r) = \frac{4}{\sqrt\pi} r^2 \biggl[c_1^2 \a_1^{3/2} e^{-\a_1 r^2}
+ c_2^2 \a_2^{3/2} e^{-\a_2 r^2} + 2c_1c_2 (\a_1\a_2)^{3/4}
e^{-\tfrac{\a_1 + \a_2}{2}r^2} \biggr],
$$
with the reassuring normalization
$\displaystyle \int_0^\infty \rho(r) \,dr = 1$.

For the energy integrals, first there are the contributions
$$
c_1^2 \biggl( \frac{3\a_1}{4} - 2Z \sqrt{\frac{\a_1}{\pi}}\,\biggr) +
c_2^2 \biggl( \frac{3\a_2}{4} - 2Z \sqrt{\frac{\a_2}{\pi}}\,\biggr).
$$
To these we add
\begin{gather*}
\int \frac{p^2}{2} P^{12} \,du 
= \biggl( \frac{\a_1\a_2}{\pi^2} \biggr)^{3/4}
\frac{3\a_1\a_2}{2(\a_1 + \a_2)}
\biggl( \frac{2\pi}{\a_1 + \a_2} \biggr)^{3/2}
= 3\sqrt{2} \frac{(\a_1\a_2)^{7/4}}{(\a_1 + \a_2)^{5/2}},
\\
\int \frac{Z}{r} P^{12} \,du 
= \biggl(\frac{\a_1\a_2}{\pi^2}\biggr)^{3/4} \frac{4\pi Z}{\a_1+\a_2}
= \frac{4Z(\a_1\a_2)^{3/4}}{\sqrt\pi(\a_1 + \a_2)},
\end{gather*}
reproducing the above result~\eqref{eq:nox-ruit} when $\a_1 = \a_2$.

Collecting our formulas, in the end we arrive at
\begin{align*}
E(\a_1,\a_2,c_1,c_2) 
&= c_1^2 \biggl( \frac{3\a_1}{4} - 2Z\sqrt{\frac{\a_1}\pi}\,\biggr)
+ c_2^2 \biggl (\frac{3\a_2}{4} - 2Z\sqrt{\frac{\a_2}\pi}\,\biggr)
\\
&\qquad + 2c_1c_2 \biggl( 3\sqrt{2} 
\frac{(\a_1\a_2)^{7/4}}{(\a_1 + \a_2)^{5/2}}
- \frac{4Z(\a_1\a_2)^{3/4}}{\sqrt\pi(\a_1 + \a_2)} \biggr).
\end{align*}
We recall that this expression is to be minimized with the constraint
$$
1 = c_1^2 + c_2^2
+ 2c_1c_2 \biggl( \frac{4\a_1\a_2}{(\a_1 + \a_2)^2} \biggr)^{3/4}.
$$
Analytically, it seems a hopeless task. Numerically, it is found that
$$
c_1 = 0.821230, \quad c_2 = 0.274403; \quad
\a_1 = 0.403059\,Z^2, \quad \a_2 = 2.664500\,Z^2.
$$
The corresponding energy is $-0.485813\,Z^2\,\mathrm{au}$, a good
estimate given the simplicity of our approach.

\vspace{6pt}

We can also now minimize with the same method
\begin{align*}
& E(\a_1,\a_2,\a_3,c_1,c_2,c_3)
\\[\jot]
&= c_1^2 \biggl( \frac{3\a_1}{4} - 2Z\sqrt{\frac{\a_1}{\pi}}\,\biggr)
+ c_2^2 \biggl( \frac{3\a_2}{4} - 2Z\sqrt{\frac{\a_2}{\pi}}\,\biggr)
+ c_3^2 \biggl( \frac{3\a_3}{4} - 2Z\sqrt{\frac{\a_3}{\pi}}\,\biggr)
\\
&\quad + 2c_1c_2 \biggl( 3\sqrt{2}
\frac{(\a_1\a_2)^{7/4}}{(\a_1 + \a_2)^{5/2}}
- \frac{4Z(\a_1\a_2)^{3/4}}{\sqrt\pi(\a_1 + \a_2)} \biggr)
+ 2c_1c_3 \biggl( 3\sqrt{2} 
\frac{(\a_1\a_3)^{7/4}}{(\a_1 + \a_3)^{5/2}}
- \frac{4Z(\a_1\a_3)^{3/4}}{\sqrt\pi(\a_1 + \a_3)} \biggr)
\\
&\quad + 2c_2c_3 \biggl( 3\sqrt{2}
\frac{(\a_2\a_3)^{7/4}}{(\a_2 + \a_3)^{5/2}}
- \frac{4Z(\a_2\a_3)^{3/4}}{\sqrt{\pi}(\a_2 + \a_3)} \biggr),
\end{align*}
constrained by
$$
1 = c_1^2 + c_2^2 + c_3^2 + 2c_1c_2 \biggl( 
\frac{4\a_1\a_2}{(\a_1 + \a_2)^2} \biggr)^{\!3/4}
+ 2c_1c_3 \biggl( \frac{4\a_1\a_3}{(\a_1 + \a_3)^2} \biggr)^{\!3/4}
+ 2c_2c_3 \biggl( \frac{4\a_2\a_3}{(\a_2 + \a_3)^2} \biggr)^{\!3/4}.
$$
Now one obtains
\begin{gather*}
c_1 = 0.647676, \quad c_2 = 0.407884; \quad c_3 = 0.070476;
\\
\a_1 = 0.302753\,Z^2, \quad \a_2 = 1.362579\,Z^2, \quad
\a_3 = 9.000725\,Z^2;
\\
\word{and} E_\mathrm{opt} = -0.496979\,Z^2\,\mathrm{au}.
\end{gather*}
We judge this good accuracy for the ground states of $\mathrm{H}$-like
ions, showing the viability of the phase space approach; the rule of
thumb ``three Gaussian type orbitals for each Slater type
orbital''~\cite{Jensen07} is fulfilled. Wigner transitions hold the
key to serious computations with Gaussian basis sets in WDFT: they
allow insight on the effects of ``negative probability'' regions for
Wigner quasidensities at low computational cost.

\smallskip

Let us come back to the case $M = 1$, for the helium series. We choose
for the quasidensity:
$$
d(\q;\p) = \frac{2}{\pi^3} e^{-\a r^2 - p^2/\a},  \word{so that}
\rho(\q) := \frac{2}{\pi^3} \int e^{-\a r^2 - p^2/\a} \,d^3p
= \frac{2\a^{3/2}}{\pi^{3/2}}\, e^{-\a r^2}.
$$
For the energy, first of all we get for the noninteracting part
\begin{equation*}
E_\mathrm{ni}(\a) = \frac{3\a}{2} - 4Z\sqrt{\frac{\a}{\pi}},
\end{equation*}
just by multiplying the result in~\eqref{eq:nox-ruit} by~2. The
minimum of just this for the neutral ion $Z = 2$ would be found at
$\a = 64/9\pi$ and is equal to $-32/3\pi$. With the assumption of a
singlet state there is no exchange energy, and the Coulomb electronic
interaction integral is easily taken care of:
\begin{align*}
\frac{1}{4} \int \frac{\rho(\q)\rho(\q')}{|\q - \q'|} \,d\q\,d\q'
= \frac{1}{4} \int
\frac{\rho(\vec R + \half\vec\rr\,)\rho(\vec R - \half\vec\rr\,)}{\rr}
\,d\vec R\,d\vec\rr = \sqrt{\frac{2\a}{\pi}}.
\end{align*}
For $Z = 2$ this is $4\sqrt{2}$ times smaller in absolute value than
the nucleus-electron energy, a smallish%
\footnote{This ratio is equal to 6.4 for the Kellner model of
$\mathrm{He}$, and is bound to be larger for the ``true'' model.}
but roughly satisfactory ratio. We thus get
$$
E_\mathrm{tot}
= \frac{3\a}{2} - (4Z - \sqrt{2}) \sqrt{\frac{\a}{\pi}}\,;
$$
the minimum is now found at $\a = (16Z^2 + 2 - 8Z\sqrt{2})/9\pi$ and
is equal to
$$
\frac{4Z\sqrt{2} - 1 - 8Z^2}{3\pi}.
$$
Thus for helium we get $\a \sim 1.53$ and $E \sim -2.3\,\mathrm{au}$: 
far above the true energy, although hardly worse than the comparable
result for hydrogen. Also, we already know from~\eqref{eq:sic-transit}
the necessary equality
$$
\int d(\q;\p;\vs,\vs)^2 \,d\q\,d\p\,d\vs 
= \int d(\q;\p)^2 \,d\q\,d\p 
= \frac{4}{\pi^6} \int e^{-2\a r^2 - 2p^2/\a} \,d\q\,d\p
= \frac{4}{(2\pi)^3}\,.
$$

What we have just done coincides with the HF calculation and
discussion of the corresponding Wigner intracules in
\cite[Sect.~7.1]{GillCONB06}, in which the trial state is exact for a
pair of uncoupled oscillators. In this reference the Wigner intracules
for noninteracting fermions in a harmonic well is computed in closed
form until $N = 8$ and they are ``\dots\ surprisingly similar to those
of qualitatively analogous atoms''. Now, in such a context it might
seem tempting to recruit to the cause the exact ground state for
\textit{interacting} harmonium, given that the intracule formula for
the interelectronic Coulomb energy is very simple:
$$
E^I_\ee(\a,k) = \int \frac{I(\vec\rr\,)}{\rr} \,d\vec\rr
= \biggl( \frac{\sqrt{\a^2 - k}}{2\pi} \biggr)^{3/2}
\int \frac{e^{-\sqrt{\a^2 - k}\,\rr^2/2}}{\rr} \,d\vec\rr
= \sqrt{\frac{2\sqrt{\a^2 - k}}{\pi}}.
$$ 
Then $E_\mathrm{tot}(\a,k) = E_\mathrm{ni}(\a,k) + E^I_\ee(\a,k)$,
where the first part of the energy has the form
$$
E_\mathrm{ni}(\a,k) = \biggl(
\frac{4\a\sqrt{\a^2 - k}}{(\a + \sqrt{\a^2 - k})^2} \biggr)^{3/2}
\Biggl( \frac{3(\a + \sqrt{\a^2 - k})}{4}
- 4Z \sqrt{\frac{2\a\sqrt{\a^2 - k}}{\pi(\a + \sqrt{\a^2 - k})}}\,
\Biggr).
$$
However, numerical calculation show only a marginal improvement in the
energy.

\subsection{Real atoms}
\label{ssc:getting-real}

For the $\H$-atom, one may consult now~\cite[Figure~5]{DahlS82} for
$F_{1s}(r,p)$, drawn from a very good set of Gaussians with $M = 10$.
The function there attains its maximum value at $(1.30,0.6)$. There is
an infinite region of damped oscillations going into negative value
regions, starting with a nodal curve going through $r = 0.5\,a_0$ at
$p$ approximately equal to~4; through $r = 1$ at $p \simeq 2.3$;
through $r = 2$ at $p \simeq 1.4$; through $r = 3$ at $p \simeq 1$;
through $r = 4$ at $p \simeq 0.8$. The amplitudes are small
($F_{1s}(1.8,1.8) = -0.0047$); but oscillations are definitively
there.

\begin{figure}[t]
\centering
\includegraphics[scale=0.8]{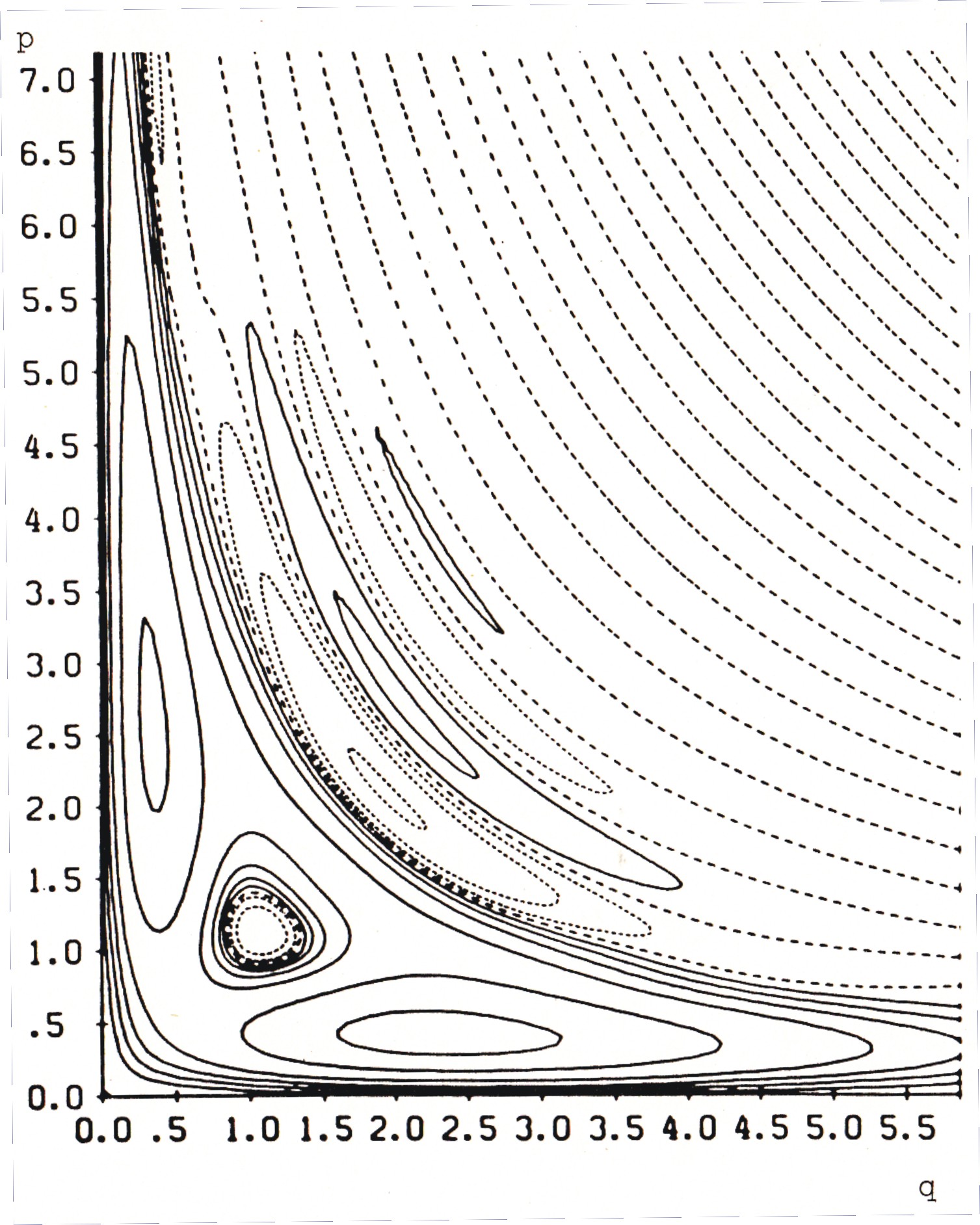}
\caption{Contour curves in hartrees for the Wigner function for the
beryllium atom. Reprinted with permission from
Ref.~\cite{SpringborgD87}.}
\label{Fig:1}
\end{figure}

Images for closed-shell atoms, based on Hartree--Fock configurations,
are given in~\cite{SpringborgD87}. Experience with atomic ground-state
Wigner functions allows one to reach the conclusion that the phase
space region supporting an atomic Wigner quasidensity separates
typically into three rough subregions. In the inner region the
function $F_{1s}$ mostly takes positive values; but it may take
negative values, due to complicated interplays among orbitals: the
one-body Wigner distribution for the ground state of neon exhibited
in~\cite{SpringborgD87} is a case in point, because of the weight of
the $2p$ orbital. One sees that in the dominant middle region, this
function takes large positive values; but negative values may also
appear, due to entanglement between electron pairs. In both these
regions the distribution barely oscillates. In the edge or ``Airy''
region we find an oscillatory decay regime. This decay of the Wigner
distribution has been rigorously proved to be generic for systems with
exponentially decaying (with linear exponents) states~\cite{Dahl95},
such as those of atoms. As it turns out, the middle region corresponds
to the region of the atom where the semiclassical approximation is
reasonable; and the nodal curve by the frontier between the middle and
the outer regions reproduces surprisingly well the border of the
occupied region in the TF model of the atom -- even for the $\H$-atom!

In the inner region the Wigner function definitely departs from the
Thomas--Fermi Ansatz, in that it is always bounded; also, as soon as
$N > 2$, one finds ``holes'' of negative quasiprobability not far from
the origin in phase space. In Figure~\ref{Fig:1} (taken
from~\cite{SpringborgD87}) the hole produced by the interference of
the $1s$ and $2s$ pairs is clearly visible.

\vspace{6pt}

On the strength of all this work we realize as well that WDFT provides
an easily visualizable bridge between the TF and quantum approaches.
We formalize this in Appendix~\ref{app:precursor}.%
\footnote{The nodal structure of the electronic Wigner function for
molecules has been investigated in~\cite{SchmiderD96}.}

\section{Natural Wigner orbitals}
\label{sec:natuerlich}

\subsection{Preliminaries}
\label{ssc:setting-table}

Now we gear up for a new approach to $\A[d]$. Generally speaking, this
is equivalent to looking for a functional~$d_2[d]$, or for a
functional~$\rho_2[d]$ (one does not need $d_2$ to compute the Coulomb
repulsion, its diagonal is enough); or even a functional $I[d]$, with
$I$ the position intracule (for which we presume there is no
representability problem), would be enough. In view of the Schmidt
theorems on best approximation \cite{Davidson76,Schmidt07}, it is
natural to use the spectral representation of $d(\q;\p;\vs;\vs')$.
Coleman's representability theorems~\cite{ColemanY00} can be construed
as implying the \textit{natural expansion}
$$
d(\q;\p;\vs,\vs') = \sum_{i=1}^\infty n_i\, f_i(\q;\p;\vs,\vs'),
$$
where the \textit{occupation numbers} $n_i$ fulfil $0 \leq n_i \leq 1$
and $\sum n_i = N$; we order them by descending size. A state with
$n_i = 0,1$ is a pinned, extremal or HF state; we already discussed
them in subsection~\ref{ssc:red-wigner}. Examples are known of
interacting systems for which $0,1$ for some of the natural orbitals
occur.%
\footnote{This point was clarified to us by the referee.}
For Coulomb systems typically the first $n_i$ for $1 \leq i \leq N$
are close to~$1$, corresponding to a state close to the best HF state;
and the others are small. A proof of infinitude of non-zero occupation
numbers in this context has been claimed in~\cite{Friesecke03}. There
are tantalizing cases, however, where the best numerical computations
stubbornly yield reduced states of finite rank --
see~\cite{NakataNEFNF01} for the first excited state of beryllium.

In the above sum, $i$ carries both spatial and spin indices. To have
pure spin eigenstates, the non-diagonal spin blocks must be zero;
moreover, in a spin-compensated, closed-shell situation the diagonal
blocks are equal. Then $N$ must be even. In this case the
\textit{spinless} quasiprobability is of the form
$$
d(\q;\p) = \sum_i \nu_i\, f_i(\q;\p),
$$
where the $f_i$ are still normalized to~$1$, but now 
$0 \leq \nu_i \leq 2$. Also, the $f_i$ verify
\begin{equation}
d \x f_i = f_i \x d = \nu_i f_i.
\label{eq:malitia-crescente} 
\end{equation}
We call these $f_i$ the natural Wigner orbitals (NWO) -- natural
Wigner spatial orbitals would be more precise. We shall also need the
natural Wigner spatial transitions, denoted $f_{ij}$. We know that in
principle they can be found from the $f_i$ and, for $d$~spinless, they
satisfy
$$
d \x f_{ij} = \nu_i f_{ij}; \qquad f_{ij} \x d = \nu_j f_{ij}.
$$

\subsection{The M\"uller functional}
\label{ssc:mulling-it-over}

For $d$ a HF state, the general form of $\A[d]$ for an electronic
interaction $f$ is well known:
\begin{align}
\A[d] &= \frac{1}{2} \biggl[\int \rho(\q_1) \rho(\q_2) f(\q_1,\q_2)
\,d\q_1\,d\q_2 - \int |\ga(1,2)|^2\, f(\q_1,\q_2) \,d1\,d2 \biggr]
\notag \\
&=: D(\rho,\rho) - X(\ga),
\label{eq:tres-tristes-tigres} 
\end{align}
where $\rho$ and $\ga$ are known functionals of~$d$; the term $X(\ga)$
is the \textit{exchange} functional. If one uses
\eqref{eq:tres-tristes-tigres} for arbitrary allowed quasidensities,
one gets a functional for the interaction energy, proceeding through
the pair density $\rho_2[d]$, denoted $\A^\HF$. Adding the kinetic and
external energy functionals, we get a functional for the total energy,
denoted $\E^\HF[d]$. Provided $f$ is positive semidefinite, its
minimum is always reached in a HF state~\cite{Lieb81PRL}. This means
that $\E^\HF[d]$ renders an \textit{upper bound} for the energy of the
system; and that Slater determinants literally do no more than scratch
the energy surface. The properties of $\E^\HF[d]$ are not very good,
besides. Clearly it is not convex. Also, it does not respect a basic
sum rule in general (this will be recalled in the next subsection).
The difference between the minimum for the total energy attained by
use of this expression and the ``true'' binding energy is generally
called the \textit{correlation energy}; it is obviously always
negative.

The problem of finding $\rho_2[d]$ has been considered in the context
of one-body density-matrix functional theory. In a remarkable paper
M\"uller in 1984~\cite{Mueller84} proposed an approximate formula for
$D_2$ amounting, in our context, in the notation
of~\eqref{eq:tres-tristes-tigres}, to the alternative functional:
\begin{equation}
\A^\mathrm{M}[d] = D(\rho,\rho) - X\bigl(\sqrt{\ga}\,\bigr).
\label{eq:mendaci-neque} 
\end{equation}
Note that because of the equivalence of operator and Moyal product
algebras, $\sqrt{\ga}$ makes sense (meaning that
$\sqrt{\ga} \x \sqrt{\ga} = \ga$), provided one treads carefully. The
concept is akin to W{\l}odarz's ``phase space wave function''
\cite{Wlodarz95,Wlodarz94}.

After a period of some obscurity, the M\"uller functional was
rediscovered~\cite{CsanyiA00,BuijseB02} and seems to be still in
fashion~\cite{FrankLSS07}. The M\"uller functional is indeed
convex~\cite{FrankLSS07}. Clearly $\A^\mathrm{M}[d] = \A^\HF[d]$ on
the subset of extremal quasidensities. However, in general
$\A^\mathrm{M}[d] \leq \A^\HF[d]$, and $\A^\mathrm{M}[d]$ actually
tends to give \textit{lower bounds} than the true values. To prove
this in general would be important.%
\footnote{See~\cite{Siedentop09} in this respect.}
It is only known for sure for $N = 1,2$ as yet~\cite{FrankLSS07}; and,
as perhaps could be expected, the proof in this last cited paper is
quite complicated. We show within this section a much simplified proof
by use of the exact functional, formulated in terms of~NWOs.

\subsection{General NWO functional theory}
\label{ssc:blasting-off}

Let the spectral representation of~$d$ be given, and denote
$$
h_i = \int f_i(\q;\p;\vs,\vs') \Bigl( \frac{p^2}{2} + V(\q\,) \Bigr)
\,d\q\,d\p \,d\vs\,d\vs'.
$$
The integrals are assumed to be finite. Then the energy is given by
$\sum_i n_i h_i + V_\ee[d_2]$. We want to express the
electron-electron repulsion $V_\ee[d_2]$ in the NWO representation.
The expansion of $d_2$ in any orthonormal basis of eigentransitions,
in particular the natural Wigner eigentransition basis, is denoted
$$
d_2(1,2) = \sum_{ijkl} D^{ij}_{kl}\, f_{ki}(1) f_{lj}(2),
$$
including spin indices.%
\footnote{\textit{Mutatis mutandis}, we borrow the notation in the
excellent reference~\cite{Piris07} here.}
Note the symmetries
\begin{equation}
D^{ij}_{kl} = {D^{kl}_{ij}}^*, \quad D^{ij}_{kl} = -D^{ji}_{kl}, \quad
D^{ij}_{kl} = -D^{ij}_{lk}, \quad D^{ij}_{kl} = D^{ji}_{lk} \,;
\label{eq:quia-absurdum} 
\end{equation}
so one can rewrite the above expansion as
$$
\sum_{i<j,\,k<l} D^{ij}_{kl} \bigl( f_{ki}(1)f_{lj}(2) 
+ f_{lj}(1)f_{ki}(2) - f_{kj}(1)f_{li}(2) - f_{li}(1)f_{kj}(2) \bigr).
$$
Notice that in this language, the HF functional corresponds to taking,
in the natural basis,
$$
D^{ij}_{kl} = \frac{n_in_j}{2} (\dl^i_k\,\dl^j_l - \dl^i_l\,\dl^j_k).
$$

In principle $d_2$ has sixteen spin blocks, but, as a consequence of
requiring pure spin states, only six differ from zero:
$$
{d_2}^{\up\up}_{\up\up}, \quad {d_2}^{\dn\dn}_{\dn\dn}, \quad
{d_2}^{\up\dn}_{\up\dn}, \quad {d_2}^{\dn\up}_{\dn\up}, \quad
{d_2}^{\up\dn}_{\dn\up}, \quad {d_2}^{\dn\up}_{\up\dn},
$$
and only three of those are independent. With this notation,
$V_\ee[d_2]$ is given by
$$
\sum_{pqrt} \braket{rt}{pq} \bigl( D^{p\up,q\up}_{r\up,t\up}
+ D^{p\up,q\dn}_{r\up,t\dn} + D^{p\dn,q\up}_{r\dn,t\up}
+ D^{p\dn,q\dn}_{r\dn,t\dn} \bigr),
$$
where
$$
\braket{rt}{pq}
:= \int \frac{f_{pr}(\q_1,\p_1)f_{qt}(\q_2,\p_2)}{|\q_1 - \q_2|}
\,d\q_1\,d\q_2 \,d\p_1\,d\p_2,
$$
with purely spatial indices. Note that the other two nonzero blocks
cannot contribute to this Coulomb integral. At this point, it is
convenient to introduce spinless density matrices by tracing:
\begin{equation}
\nu_p = n_{p\up} + n_{p\dn},  \qquad
\wt D^{pq}_{rt} := D^{p\up,q\up}_{r\up,t\up}
+ D^{p\up,q\dn}_{r\up,t\dn} + D^{p\dn,q\up}_{r\dn,t\up}
+ D^{p\dn,q\dn}_{r\dn,t\dn},
\label{eq:madre-del-cordero} 
\end{equation}
and write
$$
E = \sum_p \nu_p h_p + \sum_{pqrt} \wt D^{pq}_{rt} \braket{rt}{pq}.
$$

\smallskip

Note now that the sums:
$$
\int d_2(1,2) \,d1\,d2 
= \sum_{ijkl} D^{ij}_{kl} \int f_{ki}(1) f_{lj}(2) \,d1\,d2
= \sum_{ijkl} D^{ij}_{kl} \,\dl_i^k\dl_j^l = \sum_{ij} D^{ij}_{ij}
$$
must be $N(N - 1)/2$, the number of electron pairs, while
$$
\sum_{ijkl} \frac{n_in_j}{2}
(\dl^i_k \dl^j_l - \dl^i_l \dl^j_k \dl_i^k \dl_j^l)
= \frac{N^2}{2} - \frac{\sum n_i^2}{2} > \frac{N(N - 1)}{2},
$$
unless all $n_i$ are $0$ or~$1$, that is, the pure HF case. This is
the sum rule we alluded to in the previous subsection. Let us baptize
our ignorance
$$
\Ga^{ij}_{kl} 
:= D^{ij}_{kl} - \frac{n_in_j}{2} (\dl^i_k \dl^j_l - \dl^i_l \dl^j_k);
$$
this cumulant deserves perhaps to be called the \textit{correlation
matrix}. We now observe that, while certainly
$\int \rho(1) \rho(2) \,d1\,d2 = O(N^2)$, we have
\begin{equation}
\sum_{ijkl} \int \Ga^{ij}_{kl} f_{ki}(1) f_{lj}(2) \,d1\,d2 
= \sum_i \frac{n_i^2 - n_i}{2} = O(N).
\label{eq:que-por-diablo}  
\end{equation}
This kind of observation is useful in Thomas--Fermi theory.

\subsection{NWO for ground states of two-electron systems}
\label{ssc:gemini-module}

For two-electron atoms we can do much better. Let us invoke invariance
of the Hamiltonian under time inversion -- the latter is represented
by an antiunitary operator related to spinor conjugation that we need
not write out. This assumption is not essential, but simplifies
matters, and moreover holds in most cases of interest; it entails that
the eigenstate wave functions are real. It is most instructive to
start from a general basis of eigentransitions and construct the
natural basis out of it. This is equivalent to recasting the results
of the classic work~\cite{LowdinS56} in our language.

So let an orthonormal basis $\{f_{nm}\}$ for single-body functions on
phase space be given, arbitrary except for the properties of
Theorem~\ref{th:matrix-basis}. Consider singlet states. We want to
expand a normalized singlet 2-Wigner function $P_2 \equiv d_2$ in
terms of the~$f_{nm}$. Its spatial part must be symmetric under
exchange of coordinates; thus $P_2$ is of the form
\begin{equation}
P_2(\q_1,\q_2;\p_1,\p_2;\vs_1,\vs_2,\vs'_1,\vs'_2) =
\frac12(\up_1\dn_2 - \dn_1\up_2)
(\up_{1'}\dn_{2'} - \dn_{1'}\up_{2'})
f(\q_1,\q_2;\p_1,\p_2),
\label{eq:Macavity} 
\end{equation}
where
\begin{align}
f(\q_1,\q_2;\p_1,\p_2) = \frac{1}{4} \sum_{klmr} C_{mr} C_{kl}  
&\bigl[ f_{mk}(\q_1;\p_1) f_{rl}(\q_2;\p_2)
+ f_{rl}(\q_1;\p_1) f_{mk}(\q_2;\p_2)
\notag \\
&\quad + f_{ml}(\q_1;\p_1) f_{rk}(\q_2;\p_2)
+ f_{rk}(\q_1;\p_1) f_{ml}(\q_2;\p_2) \bigr],
\label{eq:not-there} 
\end{align}
with $C_{kl} = C_{lk}$ \textit{real} and $\sum_{kl} C_{kl}^2 = 1$.
Each group of terms is real in view of $\overline{f_{mk}} = f_{km}$
and the normalization ensures that
$$
\int P_2(\q_1,\q_2;\p_1,\p_2;\vs_1,\vs_2,\vs_1,\vs_2)
\,d\q_1\,d\q_2 \,d\p_1\,d\p_2 \,d\vs_1\,d\vs_2 = 1
$$
in~\eqref{eq:Macavity}. It is clear that in the singlet case
${d_2}^{\up\up}_{\up\up} = {d_2}^{\dn\dn}_{\dn\dn} = 0$ and that
${d_2}^{\up\dn}_{\up\dn} = {d_2}^{\dn\up}_{\dn\up} = \half f$.

The corresponding 1-Wigner distribution is obtained at once by
integration,
$$
d(\q;\p;\vs,\vs')
= (\up\up' + \dn\dn') \sum_{km} \ga_{mk} f_{mk}(\q;\p)
= \begin{pmatrix} \sum_{km} \ga_{mk} f_{mk}(\q;\p) & 0 \\[\jot]
0 & \sum_{km} \ga_{mk} f_{mk}(\q;\p) \end{pmatrix},
$$
where $\ga := [\ga_{mk}] = \bigl[ \sum_l C_{ml} C_{kl} \bigr]$ is a
symmetric (and positive semidefinite) matrix. The matrix~$C$
determines a real quadratic form that can be made diagonal by an
orthogonal change of basis $O = [O_{lm}]$,
$$
C = O\,c\,O^t, \word{implying} \ga = O\,c^2\,O^t,
$$
where $c = [c_r\,\dl_{rs}]$. These elements $c_r$ are real, but note
that some of them may be negative. Let us now make the definition:
$$
\chi_{rp} := \sum_{mk} O_{mr} f_{mk} O_{kp}, \word{so that}
f_{km} = \sum_{rp} O_{kp} \chi_{pr} O_{mr}.
$$
The diagonal $\chi_{rr} \equiv \chi_r$ are real:
$\overline{\chi_r} = \sum_{mk} O_{mr} f_{km} O_{kr} = \chi_r$, and
more generally $\overline{\chi_{rs}} = \chi_{sr}$; and it is easy to
check that
$$
\chi_i \x \chi_j 
= \sum_{mk} O_{mi} f_{mk} O_{ki} \x \sum_{ls} O_{lj} f_{ls} O_{sj} 
= \sum_{ms} O_{mi} f_{ms} O_{sj}\,\dl_{ij} = \chi_i\,\dl_{ij},
$$
and that, with $d_*$ denoting either of the two nonvanishing spin
components of~$d$,
$$
d_* = \sum_{km} \ga_{mk} f_{mk}
= \sum_{rp} \sum_{km} O_{mr} \ga_{mk} O_{kp}\,\chi_{pr}
= \sum_{rp} c^2_r\,\dl_{rp} \chi_{pr} =: \sum_r n_r \chi_r
$$
(on writing $n_r := c_r^2$),
so the $\chi_i$ solve the simultaneous equations
$$
d_* \x \chi_i =\chi_i \x d_* =  n_i \chi_i,
$$
which is~\eqref{eq:malitia-crescente} up to a factor. Furthermore,
$\sum n_i = \tr(c^2) = \tr(C^2) = 1$. The $n_i$ are the ``occupation
numbers'' for ``natural Wigner orbitals'' $\chi_i$; we have recovered
the Coleman results in this case. It remains to compute
\begin{align*}
\sum_{kl,mr} C_{mr} C_{kl}\, f_{mk}(\q_1;\p_1) f_{rl}(\q_2;\p_2) 
&= \sum_{vt,klmr} c_vc_t O_{mv} O_{rv} O_{kt} O_{lt}\,
f_{mk}(\q_1;\p_1) f_{rl}(\q_2;\p_2)
\\
&= \sum_{vt} c_v c_t\, \chi_{vt}(\q_1;\p_1) \chi_{vt}(\q_2;\p_2),
\end{align*}
and similarly for the other terms in~\eqref{eq:not-there}.

As $c_r = \pm\sqrt{n_r}$, in order to obtain $d_2[d]$ and thus $\A[d]$
one needs ``only'' to determine an infinite number of signs.%
\footnote{This amounts to an ``inversion'' of the Schmidt 
decomposition~\cite{Schmidt07} popular in entanglement theory.}
Recall that in~\eqref{eq:madre-del-cordero}, in view
of~\eqref{eq:Macavity}, the terms $D^{p\up,q\up}_{r\up,t\up}$ and
$D^{p\dn,q\dn}_{r\dn,t\dn}$ vanish. We see that
$\wt D^{pq}_{rt} = c_r c_p\,\dl_{rt}\,\dl^{pq}$ in the new basis. We
use the notations
\begin{equation}
\chi_{pr}(\q) := \int \chi_{pr}(\q,\p) \,d\p,  \qquad
L_{rp} := \braket{rr}{pp} = \int \frac{\chi_{pr}(\q_1)
\chi_{pr}(\q_2)}{|\q_1 - \q_2|} \,d\q_1\,d\q_2,
\label{eq:reality-bites} 
\end{equation}
borrowing now the eigentransitions $\chi_{pr}$ in the expansion
of~$d_2$. Note that these terms arise from correlation between
electrons with opposite spin. They are real: in view of
\eqref{eq:stoking-more-trouble}, the integrals over the momenta yield
real functions since the corresponding wave functions are real. In
particular $\chi_{pr}(\q) = \chi_{rp}(\q)$, and $L_{rp} = L_{pr}$,
too. In the weak correlation regime, when there is a \textit{dominant
state} close to the best HF~state,%
\footnote{In which precise sense close, and how close, was discussed
in~\cite{KutzelniggS68}.}
L\"owdin and Shull~\cite{LowdinS56,ShullL59} empirically found long
ago that, if conventionally $c_1$ was taken equal to~$+\sqrt{n_1}$,
then (most of) the other signs were negative. This gives rise to:
\begin{align}
d_{2\mathrm{L}}(1,2) &= \text{(spin factor) } \x \biggl[
n_1 \chi_1(\q_1) \chi_1(\q_2) - \sum_{p\geq 2} \sqrt{n_1n_p} \bigl(
\chi_{p1}(\q_1) \chi_{p1}(\q_2) + \chi_{1p}(\q_1) \chi_{1p}(\q_2)
\bigr)
\notag \\
&\qquad + \sum_{p,r\geq 2} \sqrt{n_rn_p}\, \chi_{rp}(\q_1)
\chi_{rp}(\q_2) \biggr],
\label{eq:fire-fighter} 
\\
\A^\mathrm{L}[d] &:= n_1 L_{11} - 2 \sum_{p\geq 2}
\sqrt{n_1n_p}\,L_{1p} + \sum_{p,r\geq 2} \sqrt{n_rn_p}\,L_{rp}.
\label{eq:acabaramos} 
\end{align}
In this connection the work of Kutzelnigg~\cite{Kutzelnigg63} was
decisive as well. As mentioned before, the natural orbital
construction guarantees term-by-term the most rapid approximation
to~$d_2$. The case where the sum stops merely at $p = 2$ accounts with
remarkable accuracy for a good fraction of the radial correlation
energy~\cite{LowdinS56}. We might call these the
Shull--L\"owdin--Kutzelnigg functionals (SLK functionals, for short).
For the comparison with the M\"uller functional in the next
subsection, the configuration of signs turns out to be irrelevant, and
generally we call the analogue of~\eqref{eq:acabaramos} with
\textit{any} sign choice a SLK functional.

The case for the (strict) SLK functional has been argued for
mathematically as follows. By the Rayleigh principle, the
quadratic form representing the energy has a minimum eigenvalue when
$(c_1 \ c_2 \ \cdots)^t$ is an eigenvector, that is,
$$
E_0 \begin{pmatrix} c_1 \\ c_2 \\ \vdots \end{pmatrix}
= \begin{pmatrix} 2n_1h_1 + L_{11} & L_{12} & \cdots \\
L_{12} & 2n_2h_2 + L_{22} & \cdots \\
\vdots & \vdots & \ddots \end{pmatrix}
\begin{pmatrix} c_1 \\ c_2 \\ \vdots \end{pmatrix}.
$$
Therefore, for all~$r$:
$$
E_0 = 2n_rh_r + L_{rr} + \sum_{p\neq r} L_{rp} \,\frac{c_p}{c_r}.
$$
Only the two-body part of the Hamiltonian contributes to the
off-diagonal terms. Well-known properties of the Coulomb potential
ensure that these are \textit{all positive}. Indeed, the Coulomb
potential is positive definite in that its Fourier transform is a
positive distribution. Or one can use its integral representation
\cite[Chap.~5]{LiebS10}:
$$
\frac{\pi^3}{|\q_1 - \q_2|}
= \int \frac{1}{|\vec z - \q_1|} \frac{1}{|\vec z - \q_2|} \,d^3z.
$$
Thus, whenever $n_1$ (that is to say $c_1$) is dominant, to minimize
$E_0$ we must put $c_2$ negative, and probably $c_3,c_4,\dots$ as
well. We remark that in the weak correlation regime the energy matrix
is diagonally dominant, and diagonalization of the quadratic form
proceeds without obstacle; all pivots will be negative. Diagonal
dominance ensures that the energy matrix is negative
definite~\cite{Grabowski89} as well. Nevertheless, while this
reasoning shows that there must be some negative signs, it does not
prove that all signs beyond that of~$c_1$ are negative.

A different argument to the same effect was put forward
in~\cite{GoedeckerU00}. We summarize it next. Consider the gradient of
the energy,
$$
\pd{E}{c_r} = 4h_rc_r + 2 \sum_p c_p L_{rp} - 2c_r\la,
$$
where $\la$ is a Lagrange multiplier, at the ``Hartree--Fock point'',
defined by $c_1 = 1$, $c_2 = c_3 =\cdots= 0$. For $r > 1$ only the
second term on the right hand side contributes. Thus the only way to
obtain a negative gradient is for $c_r$ to become negative in the
minimization process. Also in~\cite{GoedeckerU00} it is remarked that
for the negative hydrogen ion~$\H^-$, wherein the largest occupation
number differs significantly from~$1$, the sign rule numerically holds
true. Again, the argument is persuasive, but not conclusive, since a
moment's reflection shows that it refers to the approximation from
HF~states rather than to the exact state.

\medskip

We now list some properties and traits of the SLK functional for
singlet states.
\begin{enumerate}
\item
The sum rule $\int \rho_{2\mathrm{L}}(1,2) \,d1\,d2 :=
\int d_{2\mathrm{L}}(\q_1,\q_2;\p_1,\p_2)\, d\q_1\,d\p_1\,d\q_2\,d\p_2
= 1$ is (of course) fulfilled.
\item
The pair coincidence probability is a perfect square for any rank (and
actually any choice of signs).
\item
For pinned states, it reproduces the results of the HF functional.
\item
The correlation energy is negative.
\item
The SLK functional scales linearly:
$$
\A[d_\la] = \la\,\A[d],
$$
with $d_\la$ as in subsection~\ref{ssc:disturbare}. So the correlation
energy functional must scale linearly, too.
\item
The SLK functional satisfies known constraints from the
$D_2$-representability theory~\cite{ColemanY00}.
\end{enumerate}

The previous one is a typical closed-shell configuration, and for such
both spin components are completely identical. However, the fact of
the matter is that occupation numbers always are evenly degenerate for
any two-electron state~\cite{ColemanY00}.

\subsection{Shull--L\"owdin--Kutzelnigg versus M\"uller}
\label{ssc:SLKvsM}

Let us write the M\"uller functional \eqref{eq:mendaci-neque} in terms
of NW orbitals:
$$
\rho_{2\mathrm{M}}(\q_1,\q_2) 
:= \frac{1}{2} \sum_{jk} n_j n_k \chi_j(\q_1) \chi_k(\q_2)
- \frac{1}{2} \sum_{jk} \sqrt{n_jn_k}\,\chi_{jk}(\q_1)\chi_{kj}(\q_2).
$$
The integral over all coordinates comes out right as~1 (see below).
But clearly $d_{2\mathrm{M}}$ does not possess the antisymmetry
properties required in~\eqref{eq:quia-absurdum}. As recognized in the
original article~\cite{Mueller84}, the corresponding pair density
$\rho_{2\mathrm{M}}(\q,\q)$ can take negative values for some states.
Thus it is not surprising that the M\"uller functional tends to give
\textit{lower bounds} than the true values; when applied formally to
the hydrogen atom, this is the case. Unphysical probabilities together
with the overbinding property clearly indicate that it does not
correspond to physically realizable states.

Reference~\cite{FrankLSS07} amounts to a long analysis of the M\"uller
functional, crowned by a proof that it gives lower energy values than
the true values for helium. Now we deliver the promised simpler (as
well as more informative) proof of that theorem.

\begin{thm} 
For the isoelectronic helium-like series, the M\"uller functional
$\A^\mathrm{M}[d]$ is a lower bound to quantum mechanics.
\end{thm}

\begin{proof}
We simply compare directly the M\"uller and SLK functionals. In the
notation of \eqref{eq:reality-bites}, with 
$\rho_i(\q) = \int \chi_i(\q,\p) \,d\p$ too, the pair density coming
from the SLK functional is
\begin{gather}
\rho_{2\mathrm{L}}(\q_1,\q_2)
= \sum_i n_i \rho_i(\q_1) \rho_i(\q_2) - 2 \sum_{j\geq 2}
\sqrt{n_1n_j}\, \chi_{1j}(\q_1) \chi_{1j}(\q_2)
+ \! \sum_{i,j\geq 2,\,i\ne j} \! \sqrt{n_in_j}\,\chi_{ij}(\q_1)
\chi_{ij}(\q_2),
\notag \\
\word{normalized by}
\int \rho_{2\mathrm{L}}(\q_1,\q_2) \,d\q_1 \,d\q_2 = \sum_i n_i = 1.
\label{eq:dog-days-of-winter} 
\end{gather}
For the same object in our (two-electron, closed-shell) context, the
M\"uller functional gives
\begin{equation}
\rho_{2\mathrm{M}}(\q_1,\q_2) 
= \frac{1}{2} \rho(\q_1) \rho(\q_2) 
- \sum_i n_i \rho_i(\q_1) \rho_i(\q_2)
- \sum_{i\ne j} \sqrt{n_in_j}\, \chi_{ij}(\q_1) \chi_{ij}(\q_2).
\label{eq:manes-de-Cromwell} 
\end{equation}

It is perhaps not obvious which of the contributions
$\half \rho(\q_1)\rho(\q_2)$ or $2\sum_i n_i\rho_i(\q_1)\rho_i(\q_2)$
will be larger. However, on using
$n_i^2 = n_i - \sum_{j\neq i} n_in_j$, we compute the ``defect'':
\begin{align*}
2 \sum_i n_i \rho_i(\q_1) \rho_i(\q_2)
- \frac{1}{2} \rho(\q_1) \rho(\q_2) 
&= 2 \sum_i n_i \rho_i(\q_1) \rho_i(\q_2)
- 2 \Bigl( \sum_i n_i \rho_i(\q_1) \Bigr)
\Bigl( \sum_j n_j \rho_j(\q_2) \Bigr) 
\\
&= \sum_{j\ne i} n_i n_j \bigl( \rho_i(\q_1) - \rho_j(\q_1) \bigr)
\bigl( \rho_i(\q_2) - \rho_j(\q_2) \bigr).
\end{align*}
This must still be integrated with $1/{|\q_1 - \q_2|}$; then we see
that the positivity of the Coulomb potential alluded to in the
previous subsection does the job, establishing that
$\A^\mathrm{M}[d] < \A^\mathrm{L}[d]$.
\end{proof}

\paragraph{Remark 1}
For a chemists' chemist, the proof is surely over. A mathematically
minded one, however, would argue that we still ought to provide for
convergence of the series representing the difference in energy. To
guarantee that, consider the expression
\begin{equation}
S = \sum_{j\neq i} n_i n_j
\int \frac{F_{ij}(\q_1)F_{ij}(\q_2)}{|\q_1 - \q_2|} \,d\q_1\,d\q_2,
\label{eq:bounder-cad} 
\end{equation}
where $F_{ij} := \rho_i - \rho_j$. The Hardy--Littlewood--Sobolev
inequality \cite[Sect.~2.3]{BlanchardS96} gives, for a suitable
(sharp) constant $C_1$, the bound
\begin{align*}
S &\leq C_1 \sum_{j\neq i} n_i n_j \,\|F_{ij}\|^2_{6/5}
= C_1 \sum_{j\neq i} n_i n_j \,\|\rho_i - \rho_j\|^2_{6/5}
\\
&\leq 2 C_1 \sum_{i,j\geq 1} n_i n_j 
\bigl( \|\rho_i\|^2_{6/5} + \|\rho_j\|^2_{6/5} \bigr)
= 4 C_1 \sum_{i\geq 1} n_i \|\rho_i\|^2_{6/5}.
\end{align*}
As a marginal of a Wigner function, each $\rho_i$ is positive and lies
in $L^1(\R^3)$, with $\|\rho_i\| = 1$. However, we need to confirm
that each $\rho_i(\q)$ belongs to $L^{6/5}(\R^3)$. Finiteness of the
kinetic energy produces the estimate we need for that. As suggested
already by Moyal~\cite{Moyal49}, from
formula~\eqref{eq:stoking-trouble} for (spinless) momentum states we
obtain:
\begin{align*}
\infty > \int \frac{p^2}{2} f(\q;\p) \,d\q \,d\p
&= \frac{1}{2\pi^3} \int p^2 \ga(\p';\p'') e^{i(\q\.(\p' - \p''))}
\,\dl\bigl( \p - \half(\p' + \p'') \bigr) \,d\q \,d\p \,d\p' \,d\p''
\\
&= -\frac{1}{8} \int\bigl(\nabla_{\vecc x} - \nabla_{\vecc x'}\bigr)^2
\ga(\vecc x;\vecc x') \bigr|_{\vecc x=\vecc x'=\q}\ d\q.
\end{align*}
On account of the natural orbital expansion
$\ga(\vecc x;\vecc x') = \sum_i n_i \phi_i(\vecc x) \phi_i(\vecc x')$,
with $\rho_i = \phi_i^2$, the right hand side can be written as the
integral over~$\q$ of
$$
\frac{1}{8} \sum_{i\geq 1} n_i \biggl[
\frac{(\nabla\rho_i)^2}{\rho_i} - \nabla^2\rho_i \biggr],
$$
and its finiteness entails 
$$
\sum_i n_i \|\nabla(\sqrt{\rho_i})\|_2^2
= \sum_{i\geq 1} n_i \int \bigl( \nabla\sqrt{\rho_i} \bigr)^2 \,d\q
< \infty.
$$

The Sobolev inequality (for instance, see
\cite[Sect.~10.3]{BlanchardB92}) can now be invoked to show that
$\sum_i n_i \rho_i(\q)$ lies in $L^3(\R^3)$:
$$
\biggl\| \sum_i n_i \rho_i \biggr\|_3
\leq \sum_i n_i \|\rho_i\|_3 = \sum_i n_i \|\sqrt{\rho_i}\|_6^2
\leq C_2 \sum_i n_i \|\nabla(\sqrt{\rho_i})\|_2^2 ,
$$
for a suitable constant~$C_2$. In particular, each $\rho_i$ lies in
both $L^1(\R^3)$ and $L^3(\R^3)$, and by interpolation, it also lies
in $L^{6/5}(\R^3)$. The related H\"older inequality is
$$
\|\rho_i\|_{6/5} \leq \|\rho_i\|_1^{3/4} \|\rho_i\|_3^{1/4},
$$
and this allows us to complete the estimate of~\eqref{eq:bounder-cad}:
\begin{align*}
S &\leq 4 C_1 \sum_{i\geq 1} n_i \|\rho_i\|^2_{6/5}
\leq 4 C_1 \sum_{i\geq 1} n_i \|\rho_i\|_1^{3/2} \|\rho_i\|_3^{1/2}
\\
&\leq 2 C_1 \sum_{i\geq 1} n_i
\bigl( \|\rho_i\|_1^3 + \|\rho_i\|_3 \bigr)
= 2 C_1 \biggl( 1 + \sum_{i\geq 1} n_i \|\rho_i\|_3 \biggr),
\end{align*}
since we have already shown that the last sum is finite.

\paragraph{Remark 2}
In the literature there is at least one attempt~\cite{GritsenkoPB05}
to compare both functionals, predating~\cite{FrankLSS07}, by the way.
According to~\cite{GritsenkoPB05}, two apparent differences should
account for the M\"uller functional's tendency to overcorrelate.
\begin{enumerate}
\item
The M\"uller functional has more negative signs. Indeed, for the SLK
functional(s) it remains that some weakly occupied orbitals contribute
$+$~signs.
\item
As we have seen in our proof of the theorem, the correct expression
$\sum_j n_j \rho_j(\q_1) \rho_j(\q_2)$ is replaced by the approximate
$\half \rho(\q_1) \rho(\q_2) - \sum_j n_j \rho_j(\q_1) \rho_j(\q_2)$
-- the latter approximation being regarded as largely valid
in~\cite{GritsenkoPB05}. Note that this partition explains why the sum
rule is preserved by the M\"uller functional: on integration, the
crossed terms give zero contribution, and
$$
2 \sum_j n_j \int \rho_j(\q_1) \rho_j(\q_2) \,d\q_1\,d\q_2 = 2
= \frac{1}{2} \int \rho(\q_1) \rho(\q_2) \,d\q_1\,d\q_2.
$$
\end{enumerate}

Curiously, the extra minus signs in M\"uller's functional do not seem
to play a role.

\subsection{Harmonium test for the Shull--L\"owdin--Kutzelnigg functional}
\label{sec:harm-foretold}

To the best of our knowledge, the SLK series~\eqref{eq:fire-fighter}
and \eqref{eq:acabaramos} have never been analytically summed. Here
through the quantum phase space formalism we show that the harmonium
model provides a first example of such summation, on the way deciding
the sign dilemma for it.

The 2-representability problem is the same as for real atoms, since it
only involves the kinematics of fermions. On the other hand, the
positive definiteness property of the Coulomb potential is lost; this
and the confining nature of the one-body potential make for a peculiar
determination of signs. We ought to effect a two-step procedure:
\begin{itemize}
\item
To expand the 1-quasiprobability $d$ in natural Wigner orbitals.
\item
Then to sum the SLK series~to see (whether and) how the known
expressions for~$d_2$ and the energy are recovered. 
\end{itemize}

The difficulties are just of a technical nature. We had the
$1$-quasidensity \eqref{eq:ubi-Roma}:
\begin{equation}
d_\gs(\q,\p;\vs,\vs') = (\text{spin term}) \x 
\frac{2}{\pi^3} \, \frac{(4\om\mu)^{3/2}}{(\om + \mu)^3} \,
e^{-2(\om\mu\,q^2 + p^2)/(\om + \mu)},
\label{eq:neglecta-solent-incendia} 
\end{equation}
where now we write $\mu := \sqrt{\om^2 - k}$. Since the three
variables separate cleanly, we can drop the spin term, omit the
normalization factor, and go to the one-dimensional case
$$
d_*(u) := d_*(q,p)
:= \frac{1}{\pi} \, \frac{\sqrt{4\om\mu}}{\om + \mu} \,
e^{-2(\om\mu\,q^2 + p^2)/(\om + \mu)}.
$$
Returning to the three-dimensional $d_\gs$ is then just a matter of
notation.

{}From~\eqref{eq:agallas}, for the $2$-particle density $d_{2*}$, with
the spin term omitted and reduced from three dimensions to one, one
gleans:
\begin{align}
d_{2*}(u_1,u_2) &= \frac{1}{\pi^2} e^{-2H_R/\om} e^{-2H_r/\mu}
\label{eq:sumere-vires} 
\\
&= \frac{1}{\pi^2} \exp\biggl( - \frac{\om}{2} (q_1 + q_2)^2
- \frac{\mu}{2} (q_1 - q_2)^2 - \frac{\om^{-1}}{2} (p_1 + p_2)^2
- \frac{\mu^{-1}}{2} (p_1 - p_2)^2 \biggr).
\notag
\end{align}
Notice in passing that this is indeed a pure state, since here
$e^{-u\.Fu} = e^{-q\.Aq - p\.A^{-1}p}$, where
$$
A = \frac{1}{2} \twobytwo{\om + \mu}{\om - \mu}{\om - \mu}{\om + \mu},
\qquad
A^{-1} = \frac{1}{2} \twobytwo{\om^{-1} + \mu^{-1}}
{\om^{-1} - \mu^{-1}}{\om^{-1} - \mu^{-1}}{\om^{-1} + \mu^{-1}}.
$$
{}Formula \eqref{eq:sumere-vires} is what we need to recover. Since
mathematically \eqref{eq:neglecta-solent-incendia} is a mixed state
(in fact a maximally mixed one, as we shall see, for the given value
of the parameters), the problem is not unlike mending a broken egg.

The real quadratic form in the exponent of $d_\gs$, according
to~\cite{Titania}, must be symplectically congruent to a diagonal one,
which will turn out to be a Gibbs state. We perform the symplectic
transformation
\begin{equation}
(Q,P) := \bigl((\om\mu)^{1/4} q, (\om\mu)^{-1/4} p\bigr);
\word{or, in shorthand,}  U = Su,
\label{eq:birli-birloque} 
\end{equation}
where $S$ is evidently symplectic. Introducing as well the parameter
$\la := 2\sqrt{\om\mu}/(\om + \mu)$, the $1$-quasidensity comes from
the simple
$$
d_*(q,p) = \frac{\la}{\pi} \, e^{-\la(Q^2 + P^2)}.
$$
It is helpful to write
\begin{gather*}
\la =: \tanh \frac{\b}{2}, \word{whereby} 
\sinh \frac{\b}{2} = \frac\la{\sqrt{1 - \la^2}}
= \frac{2\sqrt{\om\mu}}{\om - \mu} \,.
\end{gather*}

{}From the well-known series formula, valid for $|t| < 1$,
$$
\sum_{n=0}^\infty L_n(x)\, e^{-x/2}\, t^n
= \frac{1}{1 - t}\, e^{-x(1+t)/2(1-t)},
$$
taking $t = -(1 - \la)/(1 + \la) = - e^{-\b}$ and
$x = 2(Q^2 + P^2)$, it follows that
$$
\frac{\la}{\pi}\, e^{-\la(Q^2 + P^2)} 
= \frac{2}{\pi}\, \sinh \frac{\b}{2} \sum_{r=0}^\infty
(-1)^r L_r(2Q^2 + 2P^2)\, e^{-(Q^2 + P^2)} e^{-(2r+1)\b/2}.
$$
All formulas involving Laguerre polynomials in this section are taken
from~\cite{Lebedev72}. We recognize the famous basis of orthogonal
oscillator Wigner eigenfunctions on phase space~\cite{Callisto},
$$
f_{rr}(Q,P)
= \frac{1}{\pi}\,(-1)^r L_r(2Q^2 + 2P^2)\, e^{-(Q^2 + P^2)}.
$$
Another well-known formula,
$$
\int_0^\infty x^\a L_m^\a(x) L_n^\a(x)\, e^{-x} \,dx
= \frac{(n + \a)!}{n!} \,\dl_{mn}, \word{for} \mathrm{Re}\,\a > 0,
$$
guarantees the correct normalization:
$\int f^2_{rr}(Q,P)\,dQ\,dP = (2\pi)^{-1}$ -- see 
Section~\ref{ssc:the-basics}.

Consequently we realize that $d_*$ is indeed a thermal bath
state~\cite{Titania}, with inverse temperature~$\b$:
$$
d_*(q,p) 
= 2\sinh \frac{\b}{2} \sum_{r=0}^\infty e^{-(2r+1)\b/2} f_{rr}(Q,P).
$$
The occupation numbers are $n_0 = 1 - e^{-\b}$ and 
$$
n_r = n_0\, e^{-r\b} = \frac{4\sqrt{\om\mu}}{\om - \mu}\, \biggl(
\frac{\sqrt\om - \sqrt\mu}{\sqrt\om + \sqrt\mu} \biggr)^{2r+1}.
$$
Clearly $\sum_r n_r = (1 - e^{-\b}) \sum_r e^{-r\b} = 1$.

\vspace{6pt}

We are now ready to recompute explicitly the SLK functional
$$
d_{2\mathrm{L}}(u_1,u_2) 
= \sum_{r,s=0}^\infty \pm \sqrt{n_r n_s}\, f_{rs}(u_1) f_{rs}(u_2).
$$
Let us write $U^2 := Q^2 + P^2$. Then, for $r \geq s$, except for
constant phase factors, the Wigner eigentransitions are 
known~\cite{Callisto}:
$$
f_{rs}(u) := \frac{1}{\pi}\,(-1)^s \sqrt{\frac{s!}{r!}} \,
(2U^2)^{(r-s)/2} e^{-i(r-s)\vth} L_s^{r-s}(2U^2) \, e^{-U^2},
$$
where $\vth := \arctan(P/Q)$. We take $f_{sr}$ to be the complex
conjugate of~$f_{rs}$.

We sum over each subdiagonal, where $r - s = l \geq 0$:
\begin{align*}
\sum_{r-s=l} & \sqrt{n_r n_s} f_{rs}(u_1) f_{rs}(u_2)
\\
&= \frac{n_0}{\pi^2}\, e^{-l\b/2} (2U_1U_2)^l e^{-il(\vth_1 + \vth_2)}
\,e^{-U_1^2-U_2^2} \sum_{s=0}^\infty \frac{s!}{(l + s)!}\, e^{-s\b}\,
L_s^l(2U_1^2) L_s^l(2U_2^2)
\\
&= \frac1{\pi^2}\, e^{-(U_1^2 + U_2^2)/\la}\, e^{-il(\vth_1 + \vth_2)}
I_l\biggl( \frac{2U_1U_2}{\sinh(\b/2)} \biggr),
\end{align*}
where $I_l$ denotes the modified Bessel function, on use of
$$
\sum_{n=0}^\infty \frac{n!}{(n + \a)!} L_n^\a(x) L_n^\a(y)\, t^n
= \frac{(xyt)^{-\a/2}}{1 - t}\, e^{-(x+y)t/(1-t)}\,
I_\a\biggl( \frac{2\sqrt{xyt}}{1 - t} \biggr).
$$
Similarly for $r - s = -l < 0$, yielding the same result replaced by
its complex conjugate. Borrowing finally the generating function
formula
$$
I_0(z) + 2 \sum_{l=1}^\infty I_l(z) \cos(l\th) = e^{z\cos\th},
$$
where, by taking $\th = \vth_1 + \vth_2 + \pi$, one obtains for the
total sum:
\begin{align*}
&\frac1{\pi^2}\, e^{-(U_1^2 + U_2^2)/\la}\,
e^{-2U_1U_2\csch(\b/2)\cos(\vth_1 + \vth_2)}
\\
&\qquad = \frac{1}{\pi^2}\, e^{-(q_1^2 + q_2^2)(\om + \mu)/2}\,
e^{-(p_1^2 + p_2^2)(\om^{-1} + \mu^{-1})/2}\, 
e^{-q_1q_2(\om - \mu)} e^{p_1p_2(\mu^{-1} - \om^{-1})},
\end{align*}
which is the correct result \eqref{eq:sumere-vires}, on the nose.

The choice $\th = \vth_1 + \vth_2 + \pi$ amounts to a clearly legal
configuration, none other than the \textit{alternating sign} rule for
the SLK functional: $c_i = (-1)^{i+1} \,\sqrt{n_i}$. Unfailingly, the
analogue of the \textit{energy functional} \eqref{eq:acabaramos} is
correctly recovered as well. Details for the latter, showing the
correct scaling behaviour~\eqref{eq:mas-sabe-el-diablo} in particular,
are given in~\cite{Pallene}.

\vspace{6pt}

We choose to recall here that extant approximations for the
exchange-correlation functional written in our terms are most often of
the following form, with an obvious generalization of our notation:
$$
\E_\mathrm{xc}[d] = - \frac{1}{2} \sum_{j,k=1} a(n_j,n_k)
\int \frac{\chi_{jk}(1) \chi_{kj}(2)}{|\q_1 - \q_2|} \,d1\,d2.
$$
These are all actually recipes for~$d_2$. For the M\"uller functional
$a(n_j,n_k) = \sqrt{n_jn_k}$. A handy list of alternatives is provided
by~\cite{Helbig06}. According to this reference, all of them (except
for the HF functional) violate antisymmetry; nearly all of them
violate the sum rule for~$d_2$; as well as invariance under exchange
of particles and holes for the correlation part. The differences
between Coulomb and confining potentials are of course considerable;
nevertheless, analytic comparison of the proposed functionals with the
exact one remains an useful exercise. This is taken up
in~\cite{Laetitia}.

Also our analysis in this section allows to throw some light from our
viewpoint~\cite{Hermione} on Gill's elusive correlation
functional~\cite{GillONB03,GillCONB06,BernardCG08}.

\vspace{6pt}

Last but not least: while our manipulations are pretty
straightforward, it would be harder to see how to go about this
problem if working on a Lagrangian hyperplane, say configuration or
momentum space, of the full phase space. (This is somewhat analogous
to the insight on the Jaynes--Cummings model brought about by the
phase space view~\cite{EiseltR89}.) The change of
variables~\eqref{eq:birli-birloque} works because there are unitary
operators -- the metaplectic representation -- effecting the
congruence and its inverse at the quantum level. We exhibit their
representatives on phase space~\cite{Littlejohn86,Ariel},
\begin{align*}
\bigl( \Xi_S^\7 \x d_* \x \Xi_S \bigr)(u) &= d_*(Su), \word{where}
\\
\Xi_S(q;p) &= \frac{2}{\sqrt{2 + (\om\mu)^{1/4} + (\om\mu)^{-1/4}}}
\exp \biggl( i \,\frac{2qp[\,(\om\mu)^{1/4} - (\om\mu)^{-1/4}]}
{2 + (\om\mu)^{1/4} + (\om\mu)^{-1/4}} \biggr);
\end{align*}
although knowledge of their existence is enough.

\appendix

\section{The Thomas--Fermi workshop}
\label{app:precursor}

\subsection{Thomas--Fermi series}
\label{aps:eppure-si-muove}
 
Many regard the TF theory of atoms as DFT \textit{avant la lettre}.
Walter Kohn himself, in his Nobel lecture~\cite{Kohn99}, dubbed his
minimum principle the \textit{formal exactification} of the TF model.
Also March has written that the ``forerunner'' Thomas--Fermi model is
completed by the Hohenberg--Kohn theorem~\cite{March95}. There is some
exaggeration in this. But the assertions do apply to our minimum
principle; that is to say, TF theory can justifiably be considered a
phase space density functional theory of the type considered in this
paper.

Let us summarize the present-day model. Combining the Pauli and
uncertainty principles, one sees that the ``radius'' of the atom goes
like $N^{-1/3}$. For a neutral atom, this together
with~\eqref{eq:basic-clue} tell us that ground state energy should be
$\propto {-}Z^{7/3}$~hartrees. The TF model then gives the value
$c_7 \equiv 0.76874512\dots$ for the proportionality constant, which
in the seventies Lieb and Simon proved (within our approximations)
exact as $Z \uparrow \infty$~\cite{LiebS77,Lieb81RMP}. Nevertheless,
traditional TF theory is often dismissed as accounting badly for the
energy, being consistently larger in absolute value. Already in 1930
Dirac introduced the exchange correction to the TF 
energy~\cite{Dirac30}.%
\footnote{It is wryly amusing to note that in this paper he wrote
down, prior to Wigner, the phase space quasiprobability associated to
the density operator. The story has been recalled in~\cite{Dahl01}: it
makes even more unsettling his reticence to accept Moyal's formulation
of quantum theory on phase space -- consult the
biography~\cite{Moyal06}.}
Although on the mark, Dirac's correction was not much used, for the
good reason that it comes with a minus sign, contributing to an ever
lower bound for the energy.

Consider quantum mechanics (of the garden or of the Wigner--Moyal
variety) for an ion with \textit{non-interacting} electrons. The
electrons would just pile up on the spectrum of states of a
hydrogenoid system, with energies $-Z^2/2n^2$, for $n$ the principal
quantum number. Regarding only closed-shell ions for simplicity, and
since such shells hold $2n^2$ electrons, we exactly have
$$
N = \sum_{n=1}^k 2n^2 = \frac{k(k + 1)(2k + 1)}3 \word{and} E = -kZ^2,
$$
for $k$ closed shells. Inverting the relation gives the following
rapidly converging series
\begin{align}
E_{\mathrm{QM},ni} &= \Bigl( -(3/2)^{1/3} Z^2 N^{1/3} + \half Z^2 -
\frac{(3/2)^{2/3}}{18} Z^2 N^{-1/3} \pm \cdots \Bigr) \,\mathrm{au}
\notag \\
&\simeq (-1.145\,Z^2 N^{1/3} + \half Z^{2} - 0.073\,Z^2 N^{-1/3}
\pm \cdots) \,\mathrm{au};
\label{eq:mirabile-dictu} 
\end{align}
The TF variational problem without Coulomb repulsion among electrons
is an exercise in~\cite[Sect.~4.1]{Thirring83}, with the result
$$
E_{\TF,ni} = -(3/2)^{1/3} Z^2 N^{1/3}.
$$
Thus the root of the problem is laid bare: the TF functional is too
rough in that it ``counts'' an infinite number of states, forbidding
to see anything but the leading term in the energy. The effect of
electron screening in such a context is to reduce $(3/2)^{1/3}$
to~$c_7$. Since the next term is independent of the number of
electrons, it ought to be the same in the screened and unscreened
theory, and that turns out to be the case: this is the Scott
correction\cite{Scott52}, attributed to inaccurate treatment of the
innermost shell, that was put on firmer ground by
Schwinger~\cite{Schwinger80}. Add Dirac's exchange correction and a
correction term of the same form, worth 2/9 of Dirac's, related to the
bulk electron kinetic energy and established by Schwinger as
well~\cite{Schwinger81}. The outcome for the neutral case is the
formula
\begin{equation}
E_\TF = (-c_7 Z^{7/3} + c_6 Z^{2} - c_5 Z^{5/3} \pm \cdots)
\,\mathrm{au},
\label{eq:Audiatur} 
\end{equation}
where $c_7$ was given above, $c_6$ is~$\half$ and $c_5 \simeq 0.2699$.
Significantly, Schwinger employed phase space arguments for all three
terms. (For the last term, his derivation can be simplified by use of
the native semiclassical approximation~\cite{Springborg84}.) Phase
space approaches encapsulate some of the non-locality which is the
bane of density gradient approximations in~DFT. A similar observation
underlies the fruitful method in~\cite{CangiLEB10}.

Not so long ago, the expression~\eqref{eq:Audiatur} was rigorously
proved to be exact to the indicated order as $Z \uparrow \infty$
\cite{FeffermanS96,CordobaFS96}. Due to the shell structure, no
continuation at $Z^{4/3}$ order exists. The correlation energy is
proportional to~$Z$ -- see~\eqref{eq:que-por-diablo} and look
up~\cite{March95} -- and so it plays little practical role.
Empirically, the result \eqref{eq:Audiatur} falls typically within
less than 0.1\% from the best Hartree--Fock values for ground-state
energies. The model is reliable for many kinds of calculations, from
diamagnetic susceptibilities to fission barriers. For a good review of
TF theory, consult~\cite{Spruch91}.

\subsection{Making sense in WDFT of the TF scheme}
\label{aps:reconsidering}

Within the old model, the energy of an atom or ion as a functional
of~$\rho$ is given by
$$
E[\rho] = V_\ext[\rho] + V_\ee[\rho] + T[\rho],
$$
where
\begin{gather}
V_\ext[\rho] = -Z \int \frac{\rho(\q)}{|\q|} \,d\q;  \qquad
V_\ee[\rho] = \frac{1}{2} \int \frac{\rho(\q)\rho(\q')}{|\q - \q'|}
\,d\q\,d\q';
\label{eq:van-dos} 
\\
T[\rho] = C_F \! \int \rho(\q)^{5/3} \,d\q,
\word{with $C_F = \dfrac{3^{5/3}\pi^{4/3}}{10}$ and the constraint}
N = \int \rho(\q) \,d\q.
\label{eq:van-tres} 
\end{gather}
Instead one can propose the functional on phase space to be
\begin{equation}
E[d] = V_\ext[d] + V_\ee[d] + T[d],
\label{eq:et-altera-pars} 
\end{equation}
where
\begin{gather*}
V_\ext[d] = -Z \int \frac{\rho(\q)}{|\q|} \,d\q, \qquad
V_\ee[d] = \frac{1}{2} \iint \frac{\rho(\q)\rho(\q')}{|\q - \q'|}
\,d\q\,d\q',  \word{as before in~\eqref{eq:van-dos};}
\\
T[d] = \frac{1}{2} \int |\p|^2 d(\q;\p) \,d\q\,d\p,
\word{with the constraint}  N = \int \rho(\q) \,d\q.
\end{gather*}
Here $\rho$ is the known functional of~$d$. Let us now compare the
functionals in~\eqref{eq:van-dos} with their partners on phase space.
The TF model in the Wigner framework simply corresponds to the purely
classical choice for~$\A[d]$. This is maybe a poor man's substitute
for~$\A[d]$; but admissible \textit{pour les besoins de la cause}. We
take issue with $T[\rho]$, obtained by semiclassical approximation --
see for instance \cite[Chap.~4]{LiebS10} or
\cite[Chap.~10]{BlanchardB92}. This is the \textit{Schwerpunkt} of the
Thomas--Fermi approach: indeed the main source of error in the TF
formalism is known to lie in the kinetic energy functional. In WDFT we
possess instead an \textit{exact} kinetic energy functional. In the
light of the work by Dahl and Springborg on atomic Wigner functions
reviewed in subsection~\ref{ssc:getting-real}, it is clear why the
uncorrected TF model fails to describe the strongly bound electrons,
and to a lesser degree the bulk electronic gas.

Precisely the minimum of $T[\rho]$ is the value taken by $T[d]$ if in
the WDFT framework one settles on the ground-state phase space density
\begin{equation}
d_\TF(\q;\p) = \frac{1}{4\pi^3} \Theta(p_F(\q) - |\p|)
:= \frac{1}{4\pi^3} \Theta \biggl(
\sqrt{\frac{2Z}{r} \chi(Z^{1/3}r/r_0)} - |\p| \biggr).
\label{eq:end-of-the-beginning} 
\end{equation}
Here $\Theta$ is the Heaviside function, $\chi$ the solution of the
famous TF equation
$\dfrac{d^2\chi(x)}{dx^2} = \dfrac{\chi(x)^{3/2}}{\sqrt{x}}$ with the
boundary condition $\chi(0) = 1$, and
$r_0 = (128/9\pi^2)^{-1/3} \simeq 0.88534$. By integration this
entails $\rho(\q) = p^3_F(\q)/3\pi^2$. For the energy, invoking the
virial theorem, from~\eqref{eq:end-of-the-beginning} by integration by
parts:
\begin{align*}
E = -\frac{1}{8\pi^3} \int \Theta\biggl( \sqrt{\frac{2Z}{r}
\chi(Z^{1/3}r/r_0)} - |\p| \biggr) p^2 \,d\p\,d\q
= \frac{8\sqrt2}{7\pi} \chi'(0) \sqrt{r_0}\,Z^{7/3},
\end{align*}
It is numerically known that $\chi'(0) \simeq -1.5881$ for the
solution we are interested in. Thus the constant before the factor
$Z^{7/3}$ is the $c_7$ reported above.

The question is: what error does \eqref{eq:end-of-the-beginning}
introduce? The essential point is that in the exact theory we are
allowed to do variations only in the \textit{narrow domain} of Wigner
quasiprobabilities, whereas the TF equation is obtained performing an
unconstrained variation (except for $\int \rho(\q) \,d\q\,d\p = N$),
not even positivity of~$\rho$ being \textit{a priori} necessary. This
explains why the TF values for the energy are grossly \textit{lower}
than the true values. For the hydrogen ground-state energy, the
uncorrected TF Ansatz gives $E_0 \simeq -0.77\,\mathrm{au}$, way
\textit{under} the $-0.5\,\mathrm{au}$ mark. Since the Coulomb
repulsion (that here should be put to zero) pulls the other way, this
is more than entirely ascribable to the use of too large a set of
phase space distributions, and to $d_\TF$ being very far from a
quantum state.%
\footnote{Recall that even approximating that ground state by the very
small set of $s$-Gaussian Wigner distributions, one obtains
$\simeq\!-0.42\,\mathrm{au}$, now of course over the mark, a pretty
better shot than the TF result.}

Said in another way, $d_\TF$ a lousy quasiprobability makes. We remark
that the associated electron density diverges at the origin as
$r^{-3/2}$, whereas a true Wigner function must be everywhere
continuous. (Also the behaviour of $d_\TF$ for large~$r$ differs from
the oscillatory exponential falloff of atomic Wigner functions
\cite{Dahl95}.) In summary, the functional is good, the domain is bad.
Or, as Parr and Yang~\cite{ParrY89} put it: ``one should try to retain
the good property of the energy functional\dots\ but to improve on the
density''.%
\footnote{Underlying this there is the motto that functional
approximation is less problematic than
representa\-bility~\cite{AyersL05}; this is why hope always
springs eternal in generalized DFT. Our method is in this spirit.}

\subsection{The formal density matrix}
\label{aps:unwinding}

Let us, like in~\cite{Mimas} and~\cite{Dahl01}, formally determine the
``reduced density matrix'' corresponding to
\eqref{eq:end-of-the-beginning}, by means of~\eqref{eq:pasodoble}.
Introducing the relative position vector $\s = \q - \q'$ and polar
coordinates for $\p$ relative to $\s$:
\begin{align}
\ga_\TF(\q;\q') 
&= \frac{1}{2\pi^2} \int_0^{p_F(R)} p^2 \,dp 
\int_0^\pi e^{ips\cos\th} \sin\th \,d\th
\nonumber \\
&= \frac{1}{\pi^2s^3} \int_0^{sp_F(R)} x\sin x \,dx =
\frac{1}{\pi^2s^3} (\sin x - x\cos x) \biggr|_0^{sp_F(R)}
\nonumber \\
&= \frac{\sin(sp_F(R))}{\pi^2s^3}
- \frac{sp_F(R)\cos(sp_F(R))}{\pi^2s^3}
\nonumber \\
&= 3\rho(R) \frac{\sin(sp_F(R)) - sp_F(R)\cos(sp_F(R))}{(sp_F(R))^3},
\label{eq:sal-si-puedes} 
\end{align}
where we have put $\vec R := (\q + \q')/2$. As a check, in view of
$$
f(x) := \sin x - x \cos x 
\sim x - \frac{x^3}{6} + \frac{x^5}{120} - x + \frac{x^3}{2}
- \frac{x^5}{24} +\cdots \sim \frac{x^3}{3} - \frac{x^5}{30} +\cdots
\word{as} x \downarrow 0,
$$
we recover $\rho(\q) = \ga_\TF(\q;\q)$. The expression is of course
ubiquitous in the theory of the noninteracting homogeneous electron
gas -- see~\cite[Chap.~6]{ParrY89} or~\cite{Eschrig96}; only its
manner of derivation is somewhat novel.

\medskip

By construction, $D_{1,\TF}$ is correctly normalized and hermitian;
however it cannot be the kernel of a density matrix and it must have
negative eigenvalues. As $Z \uparrow \infty$, the previous expression
oscillates more violently outside the vicinity of the diagonal, and
the lower eigenvalues of~$\ga_\TF$ migrate to~zero. That means that
its Wigner counterpart $d_\TF(\q;\p)$ tends to become an acceptable
quasiprobability; together with the relative vanishing of the quantum
correlations~\cite{LiebS77}, this would describe how TF theory is the
limit of the exact theory as $Z \uparrow \infty$. Better still, one
should be able to prove that $d_\TF$ approaches in some sense a true
Wigner quasidensity (it would be most interesting to see which kind of
state it looks like). We still ought to sharpen these remarks into a
new proof of the Lieb--Simon theorem. Unfortunately the appropriate
procedure to parametrize Wigner quasiprobability functions in the
limit $Z \uparrow \infty$ still eludes us; but it is difficult to
devise a conceptually simpler argument.

It also stands to reason that the Scott correction and Schwinger's
correction to the Dirac term are to be accounted for by use of the
proper Wigner quasidensities. The Dirac term itself is easily argued
for within our approach: one can read it off from
\eqref{eq:tres-tristes-tigres} and~\eqref{eq:sal-si-puedes}. We omit
this, since an identical computation is found in
\cite[Chap.~6]{ParrY89}.

\subsection{On the energy densities}
\label{aps:foresight}

We have seen in subsection~\ref{ssc:SLKvsM} that -- as suggested
already in~\cite{Moyal49} -- the natural definition for the density of
kinetic energy within phase space quantum mechanics is given by:
\begin{equation}
t_M(\q) 
= -\frac{1}{8}\bigl( \nabla_{\vecc x} - \nabla_{\vecc x'} \bigr)^2
\ga(\vecc x;\vecc x') \biggr|_{\vecc x=\vecc x'=\q}\,.
\label{eq:mas-tomate} 
\end{equation}
For the kinetic energy of a pure state given by a wavefunction
$\psi(\q)$, so that $\ga = \psi^*\psi$, this leads to:
$$
t_M(\q) = -\frac{1}{8} \nabla^2 |\psi(\q)|^2
+ \frac{1}{4} |\nabla\psi(\q)|^2,
$$
rather than the standard $\half|\nabla\psi(\q)|^2$. Remarkably, for
the ground state of a hydrogenoid ion this leads to a local form of
the virial theorem~\cite{DahlS82}. On using~\eqref{eq:storm-clouds} it
is quickly seen that
$$
t_M(\q) 
= \frac{1}{2} \biggl( \frac{Z^4}{\pi} \frac{e^{-2Zr}}{r} \biggr) 
= \frac{Z}{2r} \psi_{1s}^2(r).
$$
The above points to good local properties of quantities obtained
through the phase space formalism: a long standing tenet of the work
by Dahl and Springborg. For instance, it suggests a description of the
exchange hole close to being optimally localized~\cite{Springborg99}.
Also for many atoms apparently $t_M$ obeys a local version of the
Lieb--Oxford identities~\cite{SpringborgPS01}. Now,
$$
\nabla_{\vecc x}^2 = \quarter \nabla_{\vec R}^2 + \nabla_{\vecc r}^2
+ \nabla_{\vec R} \nabla_{\vecc r}\,; \quad
\nabla_{\vecc x'}^2 = \quarter \nabla_{\vec R}^2 + \nabla_{\vecc r}^2
- \nabla_{\vec R} \nabla_{\vecc r}\,; \quad
\nabla_{\vecc x} \nabla_{\vecc x'} = \quarter \nabla^2_{\vec R}
- \nabla^2_{\vecc r}\,.
$$
Then with $x = p_F(R)s$ and with the help of \eqref{eq:sal-si-puedes}
and~\eqref{eq:mas-tomate} we compute
$$
t_M(\q) = -\frac{3}{2} \rho(\q)
\biggl( \del_s^2 + \frac{2}{s} \del_s \biggr) f(x)\biggr|_{s=0}
= C_F{\rho(\q)}^{5/3}.
$$
We have checked that $T[d_\TF] = T[\rho_\TF]$. Now, reconsider
\eqref{eq:mas-tomate} under the more general form
$$
t(\q) = \a \nabla_{\vecc x} \nabla_{\vecc x'} \ga(\vecc x;\vecc x')
- \frac{\b}{2} \bigl(\nabla_{\vecc x}^2 + \nabla_{\vecc x'}^2\bigr)
\ga(\vecc x;\vecc x') \biggr|_{\vecc x=\vecc x'=\q}\,,
$$
where $\a + \b = \half$, in the framework of spectral expansions; it
yields
$$
t(\q) = \sum_i n_i \biggl[ \frac{(\nabla\rho_i)^2}{8\rho_i}
- \frac{\b}{2} \,\nabla^2\rho_i \biggr]
= \sum_i \frac{n_i}2 \biggl[ \bigl(\nabla\sqrt{\rho_i}\bigr)^2
- \b\,\nabla^2\rho_i \biggr].
$$
The formula with any coefficients $\a,\b$ respecting the sum rule
$\a + \b = \half$ holds \textit{only for true quantum states}. The
phase space formalism seems to ``choose'' $\a = \b = \quarter$;
suppose we had put $\a = 0$, forcing $\b = \half$, in the calculation
\eqref{eq:mas-tomate}. Then we see at once that we would have obtained
instead 
$t(\q) = -\eighth \nabla^2_{\q} \rho(\q) + C_F {\rho(\q)}^{5/3}$. This
is all very well for a real atom, since then
$\int \nabla^2_{\q}\,\rho(\q) \,d\q = 0$; but for the Thomas--Fermi
model $\int \nabla^2_{\q}\,\rho(\q) \,d\q$ diverges.

The appearance in the last display of the term of the von Weizs\"acker
type, everywhere positive and finite, is welcome. This kind of term
is known to tame the worse aspects of the TF functional, in particular
allowing for a good behaviour at the atomic nucleus site.%
\footnote{The book by Parr and Yang contains an interesting
development in this connection: it derives the improvement of the TF
energy functional by such terms from the semiclassical expansion of
the Wigner function~\cite[Sect.~6.7]{ParrY89}. However, the
coefficients obtained for the gradient corrections are not optimal,
and such expansions, when pushed to higher order, open their own box
of horrors in the form of hopelessly divergent contributions.}
Because of the Schwarz inequality, the von Weizs\"acker functional
$W[\rho]$ is convex; this means in particular that
$$
W[\rho] := \int \frac{(\nabla\rho)^2}{8\rho} \,d\q
\leq \sum_i n_i \int \frac{(\nabla\rho_i)^2}{8\rho_i} \, d\q,
$$
for $\rho_i$ as in the foregoing; for an atomic system the integral on
right hand side should precisely give the total kinetic energy.%
\footnote{Convexity of $W$ links it to relative entropy; but we
refrain from going into that.}

It is interesting to revisit harmonium in this context. It is not hard
to see~\cite{AmovilliM03} that the kinetic energy density for this
system is of the form
$$
t(q) = W\bigl[ \rho(q) \bigr]
+ 3 \rho(q)\, \frac{(\om - \mu)^2}{8(\om + \mu)}.
$$
Note that only the von Weizs\"acker term contributes in the weak
repulsion limit. We run a check on this. Enter
equation~\eqref{eq:por-viejo}; from it we glean that
$$
W[\rho_\gs](q) 
= 2\biggl( \frac{\om\mu}{\om + \mu} \biggr)^2 q^2 \rho_\gs(q).
$$
Therefore
$$
T_0 = \int W[\rho_\gs](q) \,d^3q + \frac{3(\om - \mu)^2}{4(\om + \mu)}
= \frac{3\om\mu}{\om + \mu} + \frac{3(\om - \mu)^2}{4(\om + \mu)}
= \frac{3(\om + \mu)}{4},
$$
as it should be.

\section{Characterization problem for Wigner functions}
\label{app:through-a-glass-darkly}

Nothing prevents making variational calculations with the functional
\eqref{eq:et-altera-pars}, or appropriate corrections of it, by trial
families of Wigner Gaussian orbitals, using the formulas of
subsection~\ref{ssc:Megisto}. The facts indicated at the end of
subsection~\ref{aps:eppure-si-muove} ensure reasonable results.

Those who do not fancy Gaussian basis sets then face the question of
characterizing Wigner quasiprobabilities. A one-body Wigner function
is just $N$ times a convex combination of \textit{pure} states on a
single copy of phase space. Thus the question is how to recognize a
quantum pure state representative among all functions on phase space.
Beyond reality and square summability, some necessary conditions are
easy: continuity (not smoothness, as is sometimes erroneously
assumed); the bound~\eqref{eq:bounded-joy}; positivity of the
integrals~\eqref{eq:think-positively}, and indeed of the integral over
any Lagrangian plane of the phase space. A less known necessary
condition is the following. Given a wave function $\Psi$, for each
$\q',\p'$ consider the translated wave function
$$
\Psi(\vecc x - \q')\,e^{i\p'\.\vecc x}.
$$
Its Wigner counterpart is clearly given by 
$P_\Psi(\q - \q',\p - \p')$, if $P_\Psi$ corresponds to $\Psi$. Thus
we also have the necessary condition
$$
\int M(\q;\p) P(\q - \q';\p - \p') \,d\q \,d\p \geq 0,
$$
for all pure states~$P$, all Wigner quasiprobabilities $M$ and all
$\q',\p'$.

Necessary and sufficient conditions are harder to come by. The
defining condition, namely $P = (2\pi)^{3N}(P \x P)$, is difficult to
handle. Equivalent conditions will have an oscillatory integral
somewhere; however, the following alternative may appear more
convenient to some. Performing a Fourier transform on the second set
of variables, we obtain
\begin{align*}
\wt P(\q;\r) &:= \int P(\q;\p) e^{-2i\p\.\r} \,d\p
= \frac{1}{\pi^3} \iint \Psi(\q - \z) \overline\Psi(\q + \z)
e^{2i\p\.(\z - \r)} \,d\z\,d\p
\\
&= \frac{1}{(2\pi)^3} \Psi(\q - \r) \overline\Psi(\q + \r).
\end{align*}
Thus, with $\breve P(\s) := \wt P(\s;\s)$, we get
\begin{gather*}
\breve P\bigl( \half(\q + \r) \bigr)
= \frac{1}{(2\pi)^3} \Psi(0) \overline\Psi(\q + \r); \qquad
\breve P\bigl( \half(\q - \r) \bigr)
= \frac{1}{(2\pi)^3} \Psi(q - r) \overline\Psi(0);
\\
\breve P(0) = \frac{1}{(2\pi)^3} \Psi(0) \overline{\Psi}(0);
\word{so that} \frac{\breve P\bigl( \half(\q + \r) \bigr)
\breve P\bigl( \half(\q - \r) \bigr)}{\breve P(0)} = \wt P(\q;\r).
\end{gather*}
This condition is easily seen to be sufficient as well. Also note that
$$
\Dl_q \log \wt P(\q;\r) = \Dl_r \log \wt P(\q;\r).
$$

\subsection*{Acknowledgments}
\label{ssc:in-gratitude}

We are much indebted to Jens Peder Dahl and Michael Springborg, who
over the years followed the development of the ideas presented here,
to which they have contributed numerous suggestions and needed
criticism. Many thanks are due to Daniel Gandolfo, for prompt and
knowledgeable help with numerical computations, and to Jan
Derezi\'nski for pleasurable discussions. We acknowledge the referee's
many comments and suggestions, significantly improving the paper.
JMG-B and JCV gratefully acknowledge support from the Diputaci\'on
General de Arag\'on and the Vicerrector\'ia de Investigaci\'on of the
University of Costa~Rica, respectively.

\end{document}